\documentclass[runningheads,fleqn]{cl2emult}

\usepackage{makeidx}  
\usepackage{graphicx} 
\usepackage{subeqnar} 
\usepackage{multicol} 
\usepackage{cropmark} 
\usepackage{physmu}   

\newcommand{\euler}[1]{{\usefont{U}{eur}{m}{n}#1}}

\newcommand{\umu}{\mbox{\euler{\char22}}}

\def\2eh{$2e^2/h$}  
\def\g0{$G_0$}  
  
\begin{document}
\title*{Conductance Quantisation in Metallic Point Contacts}
\toctitle{Conductance quantisation in metallic point contacts}
\titlerunning{Conductance quantisation in metallic point contacts}
\author{J.M. van Ruitenbeek}
\authorrunning{J.M. van Ruitenbeek}
\maketitle              

\index{Atomic-size contact}
\index{Conductance quantization}
\index{Mechanically controllable break junction}
\index{MCB}
\index{Point contact}
\index{Quantum wire}

\section{Introduction} 
In metallic clusters the electron wave function is confined in all three spacial dimensions, resulting in a discrete set of energy levels. In quantum wires the confinement of the electron wave function is limited to two spacial dimensions, which gives rise to the formation of a set of one-dimensional energy bands, which we refer to as modes, or channels. 

The influence of these quantum modes \index{Quantum mode}\index{Quantum channel} on the properties of metallic quantum wires has been studied experimentally by forming point contacts between metal electrodes, using scanning tunnelling microscopes, mechanically controllable break junctions,\index{Mechanically controllable break junctions} or related techniques. Where the primary observation of quantum size effects in clusters was based on cluster abundance spectra, for metal point contacts there is a somewhat analogous statistical method, which consists in recording a histogram of conductance values observed for large numbers of contacts. For simple, monovalent metals these histograms show a number of pronounced peaks close to multiples of the quantum of conductance,\index{Conductance quantum}
 $2e^2/h$. 

The interpretation of the histograms is less straightforward compared to that of the cluster abundance spectra,\index{Cluster abundance spectrum}
 since the former are the result of a combination of atomic structural features, and of the quantisation of the electron states in 1D subbands. Various experimental methods have recently been used to investigate the role of quantised modes in atomic-size contacts. These include measurements of the mechanical force on the contacts, the use of characteristic features in the current--voltage relation of superconducting contacts,\index{Superconducting contact} measurements of the thermopower,\index{Thermopower}  of the voltage-dependence of the conductance\index{Conductance fluctuations} and measurements of the shot noise\index{Shot noise} intensity. With the help of these methods it can be shown that the conductance of atomic size contact for simple metals (Au, Na, etc.) is indeed carried by a well defined set of quantum modes. However, in general the conductance is determined by the number and character of the valence orbitals of the metal atoms forming the contact. 

A very direct analogy between the physics of metal clusters and the physics of metallic point contacts becomes visible when recording conductance histograms for the alkali metals and for contacts larger than just a few atoms. A large number of peaks is observed, which have the same origin as the ``magic numbers'' \index{Magic number} in cluster abundance spectra. The observations suggest that the electronic quantum mode structure influences the mechanical stability of the nanowire, giving preference to those diameters which correspond to a filled ``shell'' of conductance modes.\index{Shell structure}

The outline of this chapter is as follows. First a brief introduction is given of the natural formalism for discussing electron transport in ballistic conductors: the Landauer theory. After introducing the experimental techniques, which are used for studying ballistic point contacts in metals, the experimental observations for the conductance of atomic-scale contacts are presented. In order to obtain a full description in terms of the quantum modes for conductance, several recently developed techniques are reviewed, which go beyond straightforward measurement of the conductance. A brief discussion is given of an unusual atomic geometry for gold contacts, which evolve into a chain of freely suspended atoms. Then shell filling effects in sodium nanowires are discussed in the context of the influence of the conductance modes on the total energy of the system. The chapter ends with an outlook on promising new developments.

\section{Landauer Theory for Ballistic Conductance}\label{sec:theory}

\begin{figure}[!t]
\includegraphics[width=.9\textwidth]{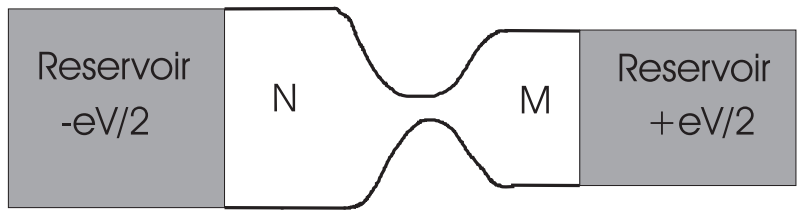}
\bigskip
\caption{Schematic representation of a ballistic point contact. The reservoirs on the left and right fully absorb the incoming electron waves. The lead on the left has a width, which admits N conductance channels, and the one on the right has M channels}
\label{fig:theory}
\end{figure}

The metallic point contacts and nanowires, which we will consider, are all of atomic size, much smaller than all characteristic scattering lengths of the system. In particular, the electron mean free path for elastic scattering on defects and impurities near the contact is assumed to be much longer than the contact size. The only elastic scattering considered is the scattering by the walls forming the boundary of the system. Also, it is assumed that the probability for scattering events, which change the spin and phase of the electron wave function, is negligible. Following the standard approach (for reviews, see \cite{beenakker,thornton,imry}) we will assume that the system can be schematically represented as in Fig.~\ref{fig:theory}. The connection between the ballistic system and the outside world is represented by electron reservoirs on each side of the contact, which are held at a potential difference $eV$ by an external voltage source. When the leads connecting the reservoirs to the contact are straight wires of constant width, there is a well defined number of conducting modes\index{Conductance mode} \index{Conductance channel} in each of these wires, say $N$ and $M$ for the left and right lead, respectively. In a free electron gas model the modes are simply plane waves, which can propagate in the current direction (to the left and right) and are standing waves in the perpendicular directions. The modes can be labelled by an index corresponding to the number of nodes in the perpendicular direction. The numbers $N$ and $M$ are limited by the requirement that the energy of the modes is lower than the Fermi energy.

The conductance of the system can now be simply expressed as \cite{beenakker,thornton,imry},
\begin{equation}  
G = {2e^2\over h} {\rm Tr}({\bf t}^{\dag}{\bf t})\; ,  \label{eq:trttdag}
\end{equation}  
where $e$ is the electron charge, $h$ is Planck's constant, and ${\bf t}$ is an $N\times M$ matrix with matrix element $t_{mn}$ giving the probability amplitude for an electron wave in mode $n$ on the left to be transmitted into mode $m$ on the right of the contact. It can be shown that the product matrix ${\bf t}^{\dag}{\bf t}$ can always be diagonalised by going over to a new basis, consisting of linear combinations of the original modes in the leads. Further, the number, $N_{\rm c}$, of non-zero diagonal elements is only determined by the number of modes at the narrowest cross section of the conductor \cite{brandbyge97,cuevas98}.  Equation~(\ref{eq:trttdag}) thus simplifies to
\begin{equation}
G = {2e^2\over h} \sum_{n=1}^{N_{\rm c}} T_n \; ,  \label{eq:landauer}
\end{equation}
where $T_n = |t_{nn}|^2$ and the index refers to the new basis.

Under favourable circumstances, all transmission probabilities,\index{Transmission probability} $T_n$, can be close to unity. As an example, for a smooth (``adiabatic'') and long wire the modes at the narrowest cross section couple exclusively to a single mode at either side of the contact, and the expression for the conductance further simplifies to
\begin{equation}
G = N_{\rm c} G_0\;  ,  \label{eq:CQ}
\end{equation}
where $G_0 = 2e^2/h$ is the conductance quantum\index{Conductance quantum}. For increasing diameter of the contact the number $N_{\rm c}$ increases each time a new mode fits into the narrowest cross section. This number is limited by the requirement that the kinetic energy for motion in the perpendicular direction is smaller than the Fermi energy. For a 2-dimensional (2D) system this can be expressed as $(\hbar^2/2 m)(\pi N_{\rm c}/W)^2 < E_{\rm F}$, with $W$ the width of the contact, which leads to $N_{\rm c} = {\rm Int}(2W/\lambda_{\rm F})$, with $\lambda_{\rm F}$ the Fermi wavelength. For a 3D metallic contact $N_{\rm c} \approx (\pi R/\lambda_{\rm F})^2$, with $R$ the contact radius. This quantisation of the conductance\index{Conductance quantization} was first observed in experiments on 2D electron gas devices by van Wees et al.\   \cite{wees} and by Wharam et al.\   \cite{wharam}, where $\lambda_{\rm F}\simeq 400$~\AA\  is much larger than the atomic scale. 

\section{Experimental Techniques}
The experimental tools for fabricating atomic-scale contacts are mostly based on a piezoelectric actuator for the adjustment of the contact size between two metal electrodes. Standard Scanning Tunnelling Microscopes\index{Scanning tunnelling microscope}  (STM)\index{STM} are often used for this purpose \cite{agrait1,pascual1,olesen}. The tip of the STM is driven into the surface and the conductance is recorded while gradually breaking the contact by retracting the tip. The first experiment of this type was reported by Gimzewski and M\"oller \cite{gimzewski}.

\begin{figure}[!t]
\includegraphics[width=.8\textwidth]{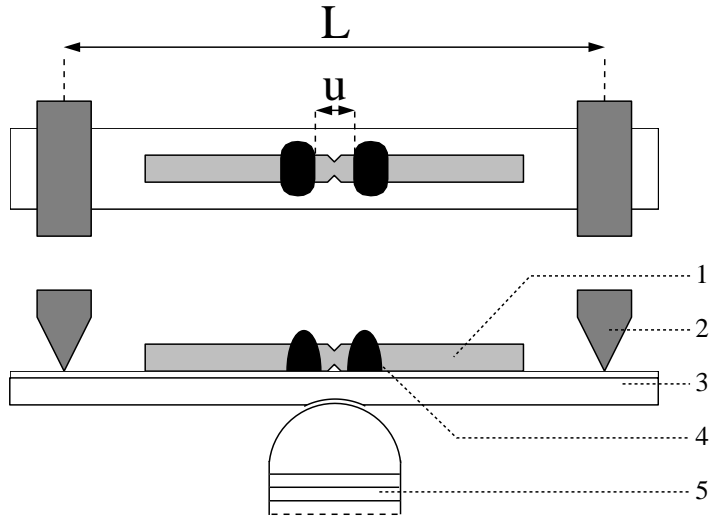}
\caption{ Schematic top and side view of the mounting of a MCB, where the metal to be studied has the form of a notched wire (1), which is fixed onto an insulated elastic substrate (3) with two drops of epoxy adhesive (4) very close to either side of the notch. The substrate is mounted in a three-point bending configuration between the top of a stacked piezo element (5) and two fixed counter supports (2). This setup is mounted inside a vacuum can and cooled down to liquid helium temperatures. Then the substrate is bent by moving the piezo element forward. The bending causes the top surface of the substrate to expand and the wire to break at the notch. Typical sizes are $L\simeq 20$~mm and $u\simeq 0.1$~mm  }
\label{fig:MCB} 
\end{figure}
A practical tool for the purpose of studying metallic quantum point contacts is the Mechanically Controllable Break-junction\index{Mechanically controllable break junction} (MCB)\index{MCB} technique \cite{muller92a}. The principle is illustrated in Fig.~\ref{fig:MCB}.
By breaking the metal, two clean fracture surfaces are exposed, which remain clean due to the cryo-pumping action of the low-temperature vacuum can. This method circumvents the problem of surface contamination of tip and sample in STM experiments, where a UHV chamber with surface preparation and analysis facilities are required to obtain similar conditions. The fracture surfaces can be brought back into contact by relaxing the force on the elastic substrate, while a piezoelectric element is used for fine control. The roughness of the fracture surfaces results in a first contact at one point, and experiments usually give no evidence of multiple contacts. In addition to a clean surface, a second advantage of the method is the stability of the two electrodes with respect to each other. From the noise in the current in the tunnelling regime one obtains an estimate of the vibration amplitude of the vacuum distance, which is typically less than $10^{-3}$~\AA. The stability results from the reduction of the mechanical loop which connects one contact side to the other, from centimetres, in the case of an STM scanner, to $\sim 0.1$~mm in the MCB.  

Conductance properties of atomic-size contacts can even be studied by still simpler methods. When breaking the contact between two regular wires under ambient conditions \cite{costa} or switching a commercial relay \cite{hansen} and when recording the evolution of the conductance with a time resolution of the order of microseconds, the conductance can be observed to decrease by atomic steps during the last stages of the contact breaking. Such methods allow rapid accumulation of histograms of conductance values for a large number of such scans.

\section{Conductance of Atomic-Scale Contacts} 

\begin{figure}[!b]
\includegraphics[width=.8\textwidth]{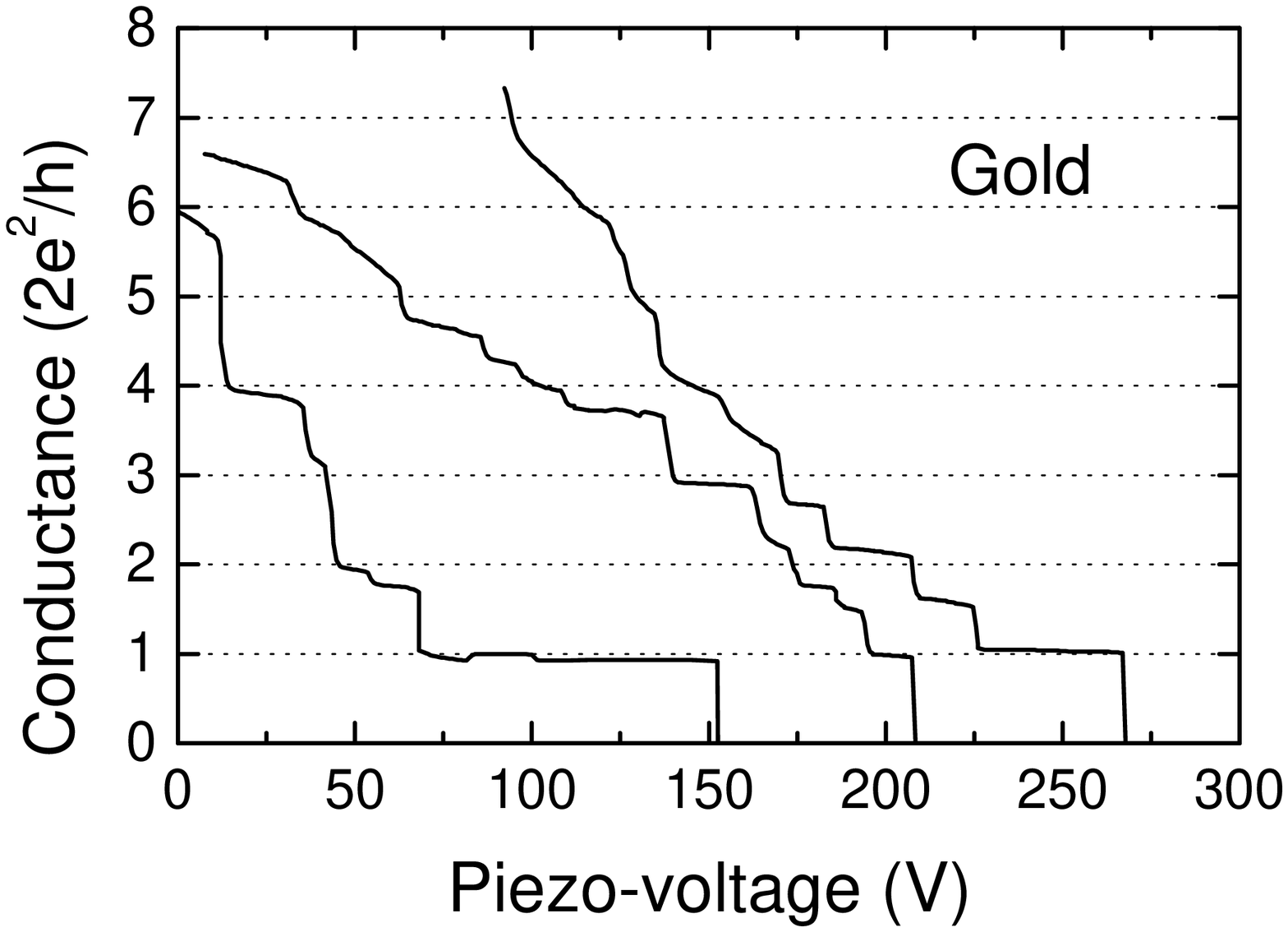}
\caption{Three typical recordings of the conductance $G$ measured in atomic size contacts for gold\index{Au} at helium temperatures, using the MCB technique. The electrodes are pulled apart by increasing the piezo-voltage. The corresponding displacement is about 0.1~nm per 25~V. After each recording the electrodes are pushed firmly together, and each trace has new structure. (After J.M. Krans \protect\cite{krans96b})  }
\label{fig:gold} 
\end{figure}

\begin{figure}[!t]
\includegraphics[width=.8\textwidth]{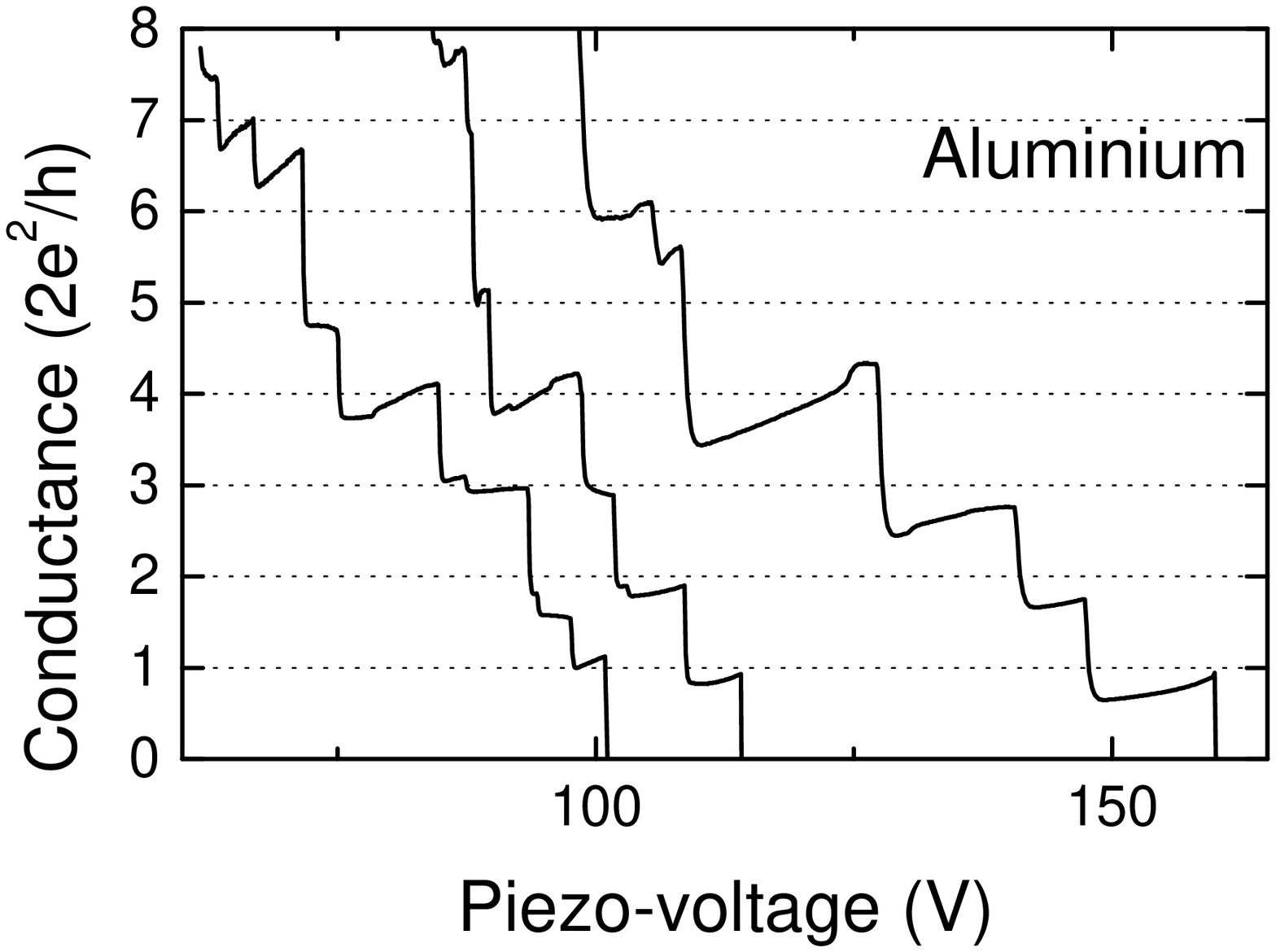}
\caption{Three examples  of the conductance measured in atomic size contacts for an aluminium\index{Al} MCB junction at 4.2~K, as a function of the piezo-voltage.  (After \protect\cite{krans93})  }
\label{fig:aluminium} 
\end{figure}

\begin{figure}[!t]
\includegraphics[width=.8\textwidth]{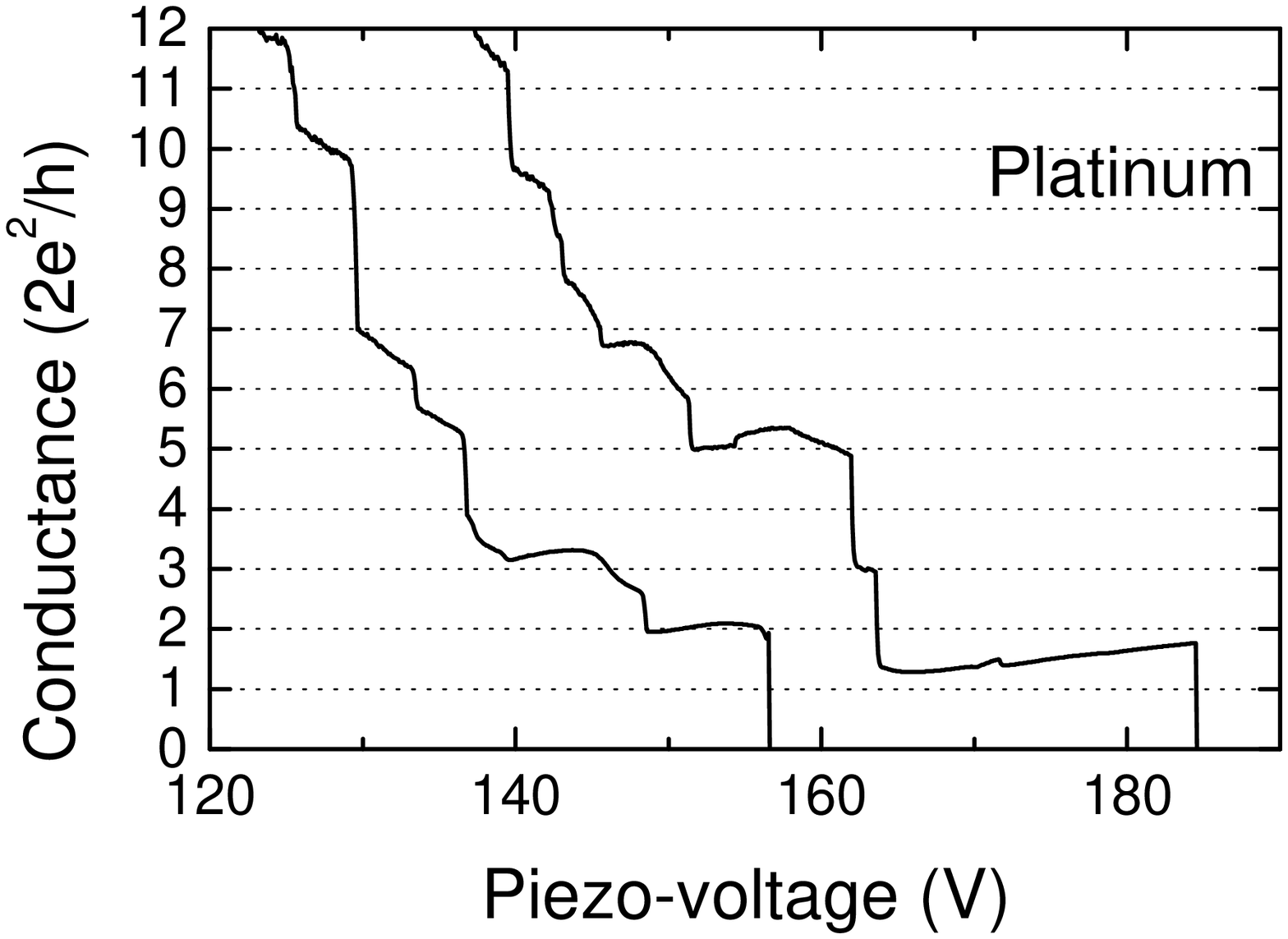}
\caption{The conductance for a platinum\index{Pt} junction at 1.3~K as a function of piezo-voltage for two successive scans. (After  \protect\cite{krans93})}
\label{fig:platinum} 
\end{figure}

Figure~\ref{fig:gold} shows some examples of the conductance measured during breaking of a gold\index{Au} contact at low temperatures, using an MCB device. The conductance decreases by sudden jumps, separated by plateaux, which have a negative slope, the higher conductance the steeper. Some of the plateaux are remarkably close to multiples of the conductance quantum, $G_0$; in particular the last plateau before loosing contact is nearly flat and very close to 1~$G_0$. Closer inspection, however, shows that many plateaux cannot be identified with integer multiples of the quantum unit, and the structure of the steps is different for each new recording. Also, the height of the steps is of the order of the quantum unit, but can vary by more than a factor of 2, where both smaller and larger steps are found. Drawing a figure such as Fig.~\ref{fig:gold}, with grid lines at multiples of $G_0$, guides the eye to the coincidences and may convey that the origin of the steps is in quantisation of the conductance.\index{Conductance quantization} However, in evaluating the graphs, one should be aware that a plateau cannot be farther away than one half from an integer value, and that a more objective analysis is required. Still, it is clear that we can use these fairly simple techniques to produce and study atomic-scale conductors, for which the conductance is dominated by quantum effects. The interpretation of graphs as in Fig.~\ref{fig:gold} will be the subject of this and the following sections. 

The fact that the atomic structure and orbital character of the electron modes is important for an interpretation of the conductance can already be deduced by comparing the conductance curves for various metals. Figure~\ref{fig:aluminium} shows three examples of conductance curves for aluminium\index{Al} atomic-size contacts, and  Fig.~\ref{fig:platinum} shows results for platinum.\index{Pt} In both cases, in particular for aluminium, we find that many plateaux have an anomalous slope: the conductance {\it increases} when pulling the contact, in contrast to the results for gold. For aluminium, the last plateau before breaking is still close to one unit of conductance, but one frequently observes the conductance diving below this value, and then recovering to nearly 1~$G_0$, before contact is lost. Platinum, on the other hand, has a last conductance value, which is usually of order two times larger and also the size of the jumps is somewhat larger.

For all metals the transition between the plateaux is very sudden and sharp. Such sudden transitions would not be expected in a model which describes the conductance in terms of a set of conduction modes, which are gradually pinched-off by reducing the contact diameter. Indeed, the jumps find their origin in sudden rearrangements of the atomic structure of the contact \cite{todorov93,muller92b}. Upon stretching of the contact, the stress accumulates elastic energy in the atomic bonds over the length of a plateau. This energy is suddenly released in a transition to a new atomic configuration, which will typically have a smaller contact size. Such atomic-scale mechanical processes were first described by Sutton and Pethica \cite{sutton90} and by Landman et al.\   \cite{landman}. 

\begin{figure}[!t]
\includegraphics[width=.7\textwidth]{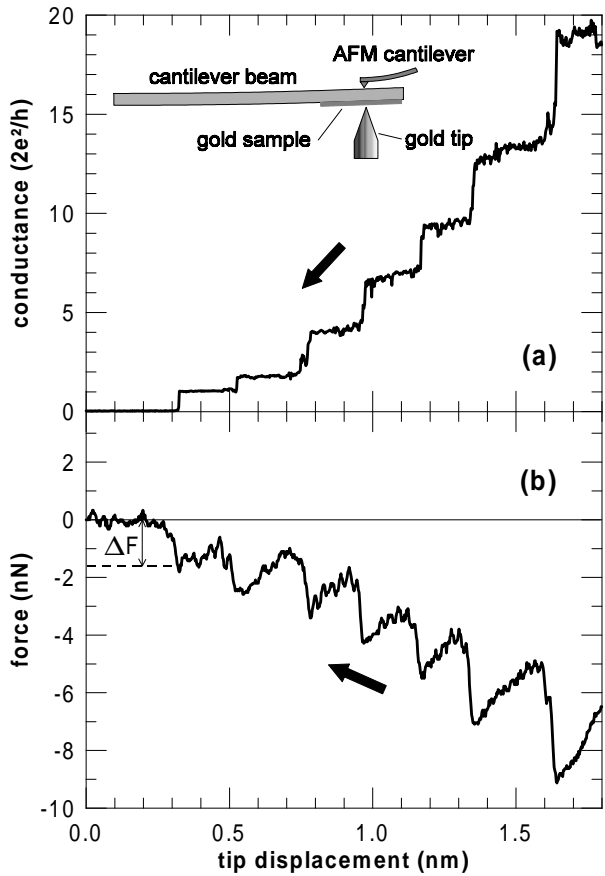}
\caption{Simultaneous measurement of force\index{Quantum force}
 and conductance on atom scale point contacts for Au.\index{Au} The sample is mounted on a cantilever beam and the force between tip and sample is measured by the deflection of the beam 
using an Atomic Force Microscope (AFM). The measurements are done in air at room temperature. (From \protect\cite{rubio}) }
\label{fig:rubio}
\end{figure}

\begin{figure}[!t]
\includegraphics[width=.6\textwidth]{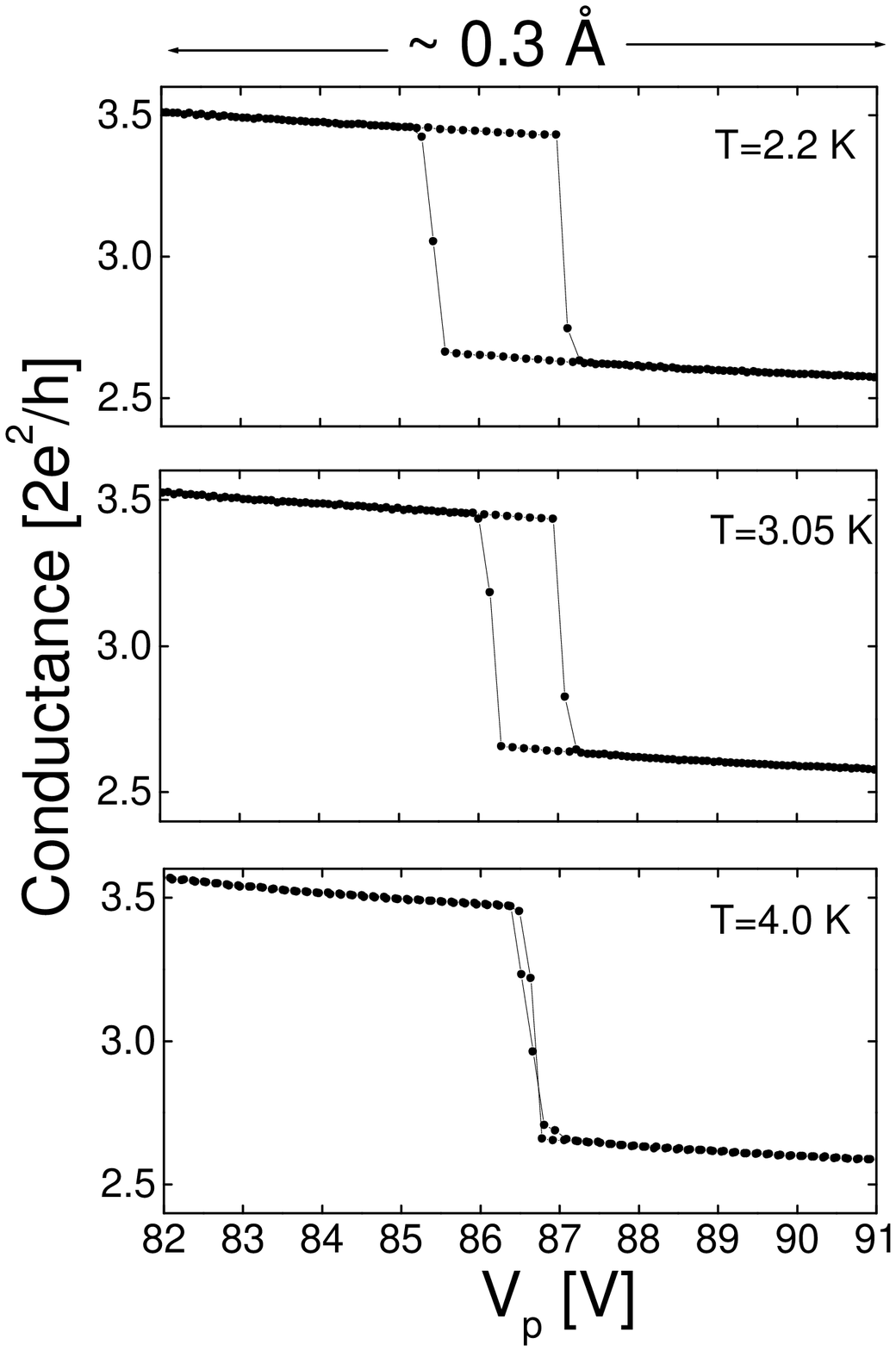}
\caption{Measurements of the conductance for an atomic-size gold\index{Au}  contact while sweeping the piezo-voltage forward and back over a single step. The curves have been recorded at three different temperatures while maintaining the sweep centred around the same jump. Clear hysteresis of the order of 0.1~\AA\  is observed at 2.2~K. At 3.1~K the hysteresis is half as large, and at 4.2~K it has disappeared. (After J.M. Krans \protect\cite{krans96a}) }
\label{fig:hyst}
\end{figure}

The first direct proof for atomic rearrangements at conductance steps was provided in an experiment by Rubio, Agra\"{\i}t and Vieira \cite{rubio}, where the conductance for atomic-size gold contacts was measured simultaneously with the force on the contacts (Fig.~\ref{fig:rubio}). The stress accumulation on the plateaux and the coincidence of the stress relief events with the jumps in the conductance can be clearly distinguished. Presently, the experiment has only been reported for gold\index{Au} at room temperature. At low temperatures, the evidence for the atomic structure related nature of the jumps comes from analysis of the dynamic behaviour of the jumps. Generally, hysteresis is observed in the position of the jumps when retracing the curve immediately after a jump is found (Fig.~\ref{fig:hyst}). When increasing the bath temperature or the current through the contact (which indirectly heats the contact) the width of the hysteresis is seen to gradually decrease until it is reduced to zero, and spontaneous jumps between the two conductance values are observed, which have a thermally activated behaviour \cite{muller92b,krans96a,brom}. These observations find a natural interpretation in terms of jumps between two distinct atomic configurations for the contact, separated by an energy barrier. Various recent molecular dynamics simulations confirm this scenario \cite{brandbyge97,todorov96,landman96,mehrez1,mehrez2}. 

\begin{figure}[!t]
\includegraphics[height=.8\textwidth,angle=270]{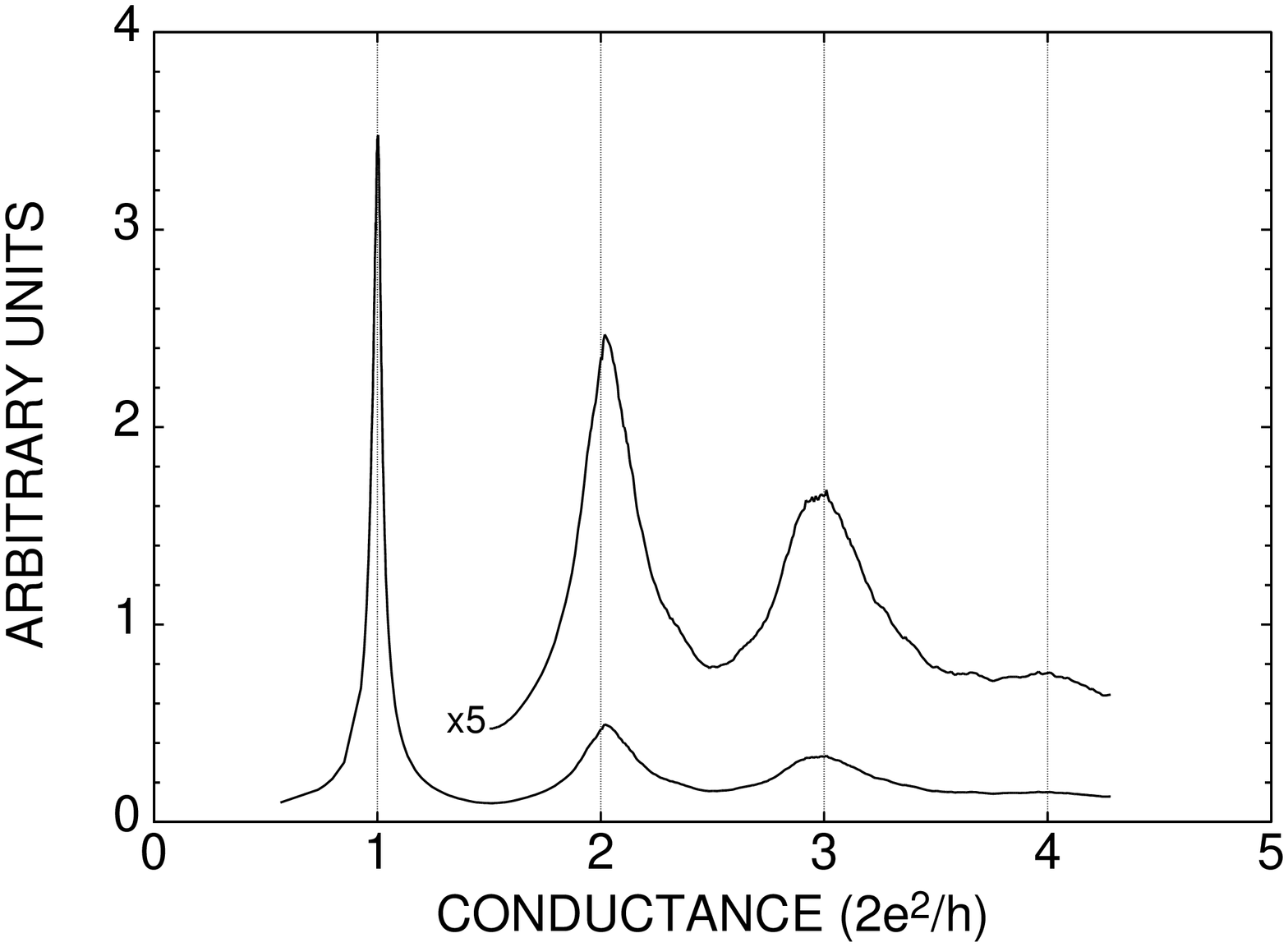}
\caption{Histogram representing the relative weight which each conductance value has in the experiments. The histogram is constructed from 227 conductance curves recorded while breaking Au\index{Au}  contacts, using an STM under UHV at room temperature.  (From \protect\cite{brandbyge95})}
\label{fig:histAu}
\end{figure}

\section{Histograms of Conductance Values}
\begin{figure}[!t]
\includegraphics[width=.8\textwidth]{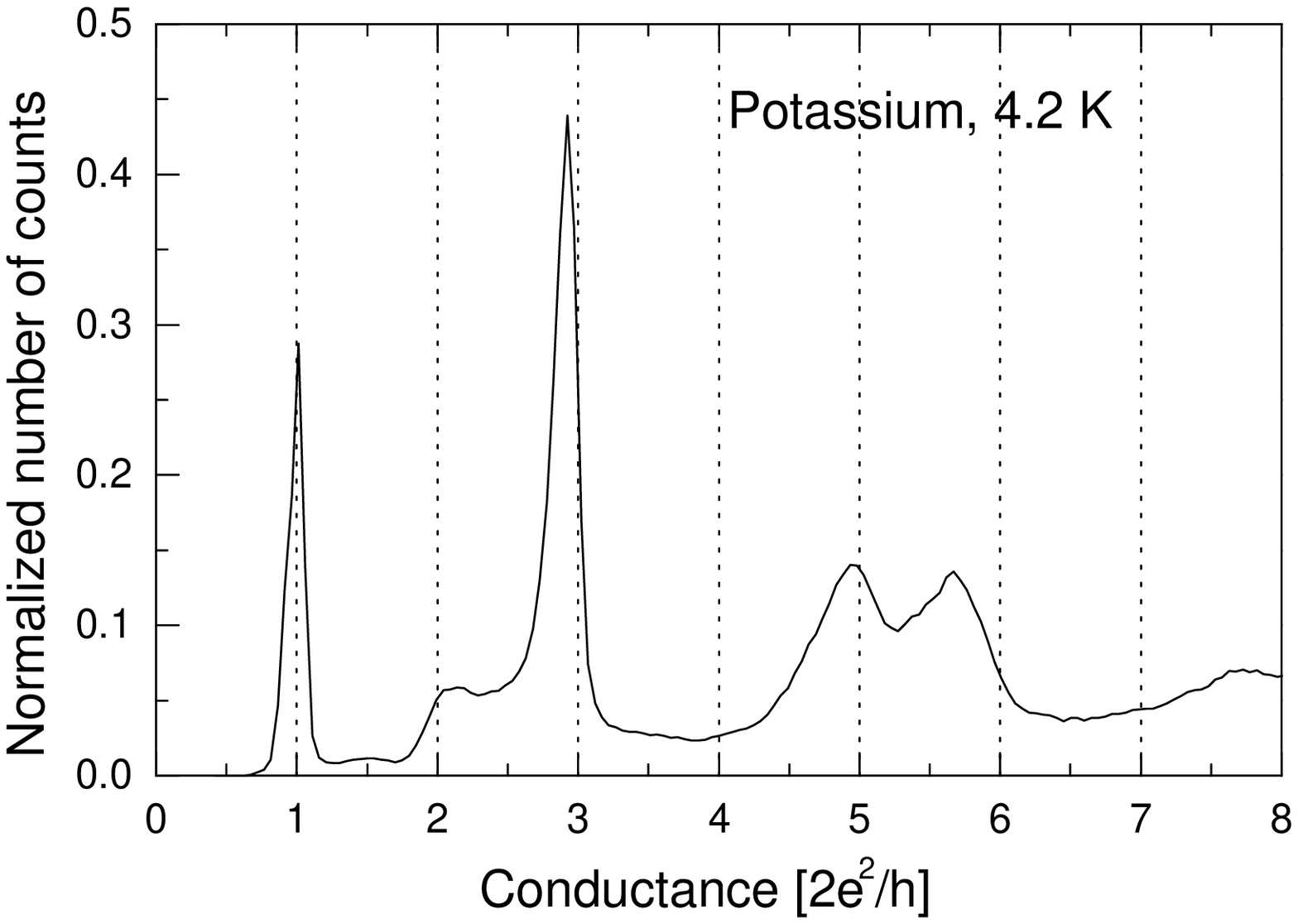}
\caption{ Histogram of conductance values, constructed from 
$G(V_{\rm p})$-curves measured for potassium\index{K} at 4.2~K with an MCB device, involving several thousand individual measurements. The characteristic sequence of peaks ($G= 1,3,5,6$) is regarded as a signature for conductance quantisation. (From \protect\cite{yanson98b})}
\label{fig:histK}%
\end{figure}

\index{Conductance histogram}
The atomic configuration of a contact adjusts itself in response to the externally applied stress, and evolves depending on the starting configuration of the contact at larger size. The fact that each conductance curve differs in many details from previous curves reflects the fact that the atomic configuration of the contact for given conductance is different in each run. However, as observed in the previous section, there appears to be a certain preference for conductance values near integer multiples of the quantum unit. For gold near 1~$G_0$ this is immediately obvious from the examples in Fig.~\ref{fig:gold}. A general and objective method of analysis was introduced \cite{olesen95,krans95}, which consists in recording histograms of conductance values encountered in a large number of runs. Figure \ref{fig:histAu} shows a histogram for gold\index{Au}  measured using a room temperature STM under UHV conditions \cite{brandbyge95}. Up to four peaks are found centred near the first four multiples of $G_0$. For sodium\index{Na} in a low temperature experiment using the MCB technique, a histogram with peaks near 1, 3, 5 and 6 times $G_0$ was observed \cite{krans95}. Figure \ref{fig:histK} shows a similar result for potassium.\index{K} The fact that peaks near 2 and 4~$G_0$ are absent points at an interpretation in terms of a smooth, near-perfect cylindrical symmetry of the sodium contacts. Sodium indeed forms a very good approximation to a free electron system, and the weakly bound $s$-electrons strongly reduce surface corrugation. This is also why the best abundance spectra\index{Cluster abundance spectrum} for clusters, with pronounced features at magic numbers\index{Magic number} have been obtained for the alkali metals \cite{de Heer}. For a model smooth, cylindrically symmetric contact with continuously adjustable contact diameter 
\cite{bogachek90,torres94}, the conductance increases from zero to 1~$G_0$ as soon as the diameter is large enough, so that the first conductance mode is occupied. When increasing the diameter further, the conductance increases by two units because the second and third modes are degenerate. The modes are described by Bessel functions (assuming a hard wall boundary potential) and the first mode is given by the $m=0$ Bessel function, which is not degenerate. The second and third modes are the degenerate $m=\pm 1$ modes, followed by $m=\pm 2$ for further increasing contact diameter. The next mode which will be occupied corresponds to the second zero of the $m=0$ Bessel function, and is again {\it not} degenerate. Thus the conductance for such a contact should increase by 1, 2, 2 and 1 units, producing just the series of conductance values observed in the sodium experiment. The slight shift of the peaks in Fig.~\ref{fig:histK} below the integer values can be attributed to an effective series resistance due to back-scattering on defects near the contact. Model simulations 
\cite{torres96} of the histogram are in close agreement with the experiment, including the shift of the peaks. The fact that gold histograms show peaks at all the first four quantum values may be explained by a stronger deviation from cylindrical symmetry in gold, which then lifts the degeneracy of the modes. Model simulations are indeed able to reproduce the shape of the gold histograms \cite{brandbyge95}. 

\begin{figure}[!b]
\includegraphics[height=.8\textwidth,angle=270]{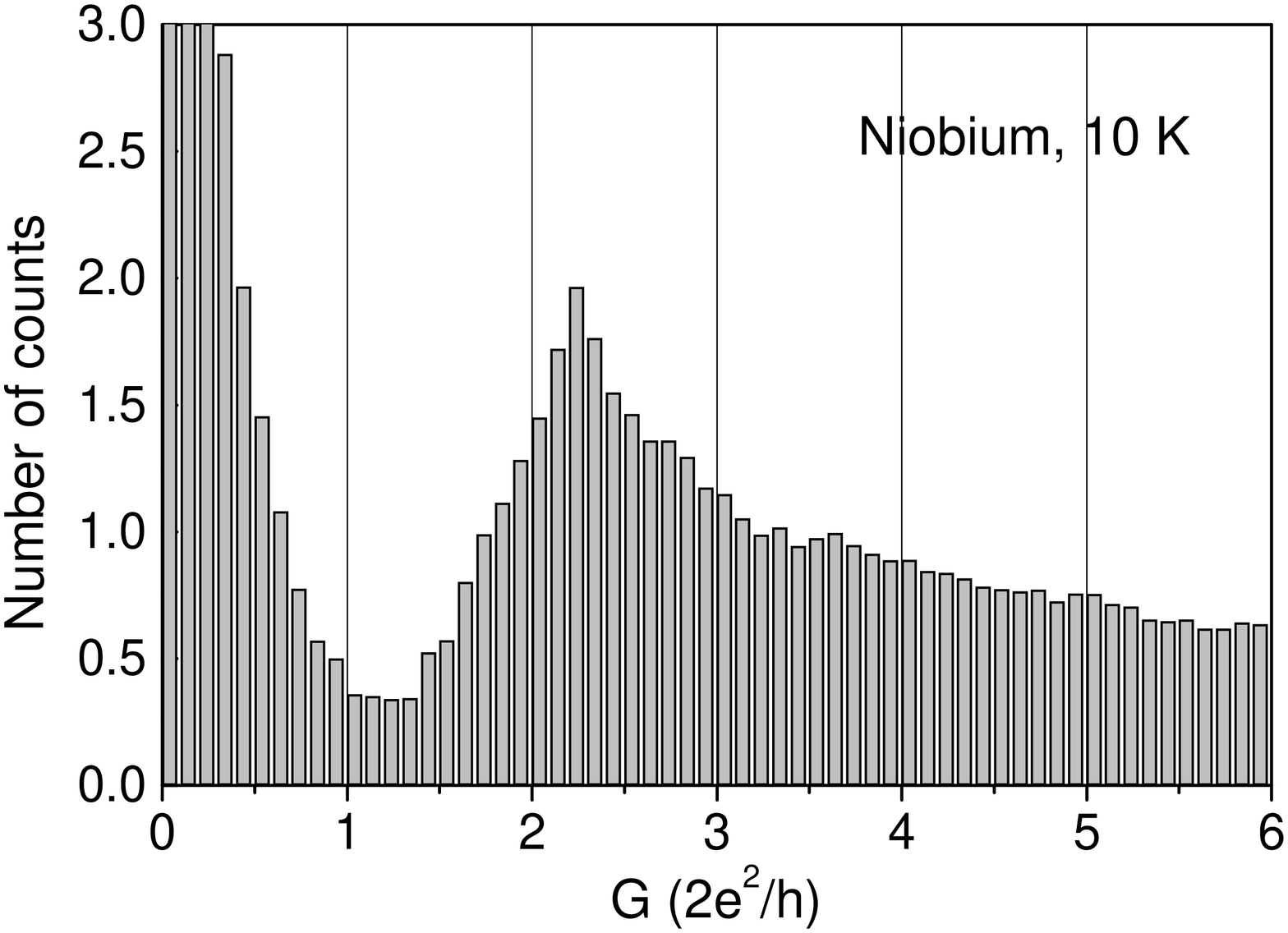}
\caption{Histogram constructed from 2400 individual conductance curves for a niobium\index{Nb} sample. Each curve was recorded while stretching the contact to break, using the MCB technique at a temperature of 10~K, which is just above the superconducting transition temperature. The conductance was measured using a DC voltage bias of 20~mV. (From B. Ludoph \protect\cite{ludoph98})}
\label{fig:histNb}
\end{figure}

Not all metals show such pronounced histogram peaks near integer conductance values. The most clear-cut results are obtained only for monovalent metals. The alkali metals Li, Na and K show a histogram structure as represented by Fig.~\ref{fig:histK}, although for Li the shift of the peaks is somewhat stronger \cite{yanson98b}. The noble metals Cu, Ag and Au show histograms as in Fig.~\ref{fig:histAu} \cite{hansen} . The details such as the shift in position, the width and relative height of the peaks can be different depending on the experimental conditions \cite{hansen,brandbyge95,krans95,gai,costa97a,costa97b}.  Most other metals only show a rather broad first peak, which reflects the conductance of a single atom contact (see below). This peak can generally not be identified with an integer value of the conductance; for example niobium\index{Nb} shows a wide peak centred near 2.5~$G_0$ (Fig.~\ref{fig:histNb}) and similar results have been obtained for Pb\index{Pb} \cite{ludoph98}. On the other hand there are a few examples of multivalent metals, which show pronounced peaks in the histograms, among which aluminium\index{Al} \cite{yanson97}. As we shall discuss below, the histogram for Al throws doubt upon a straightforward interpretation of the histogram peaks in terms of conductance quantisation. For some systems other than simple metals evidence has been obtained for preferred conductance values near multiples of the quantum of conductance (carbon nanotubes, semimetals, metal oxides, etc.). These will not be discussed here, since the nature of the contact and the mechanism of contact formation is expected to be very different from that in ordinary metals.

\begin{figure}[!b]
\includegraphics[width=.8\textwidth]{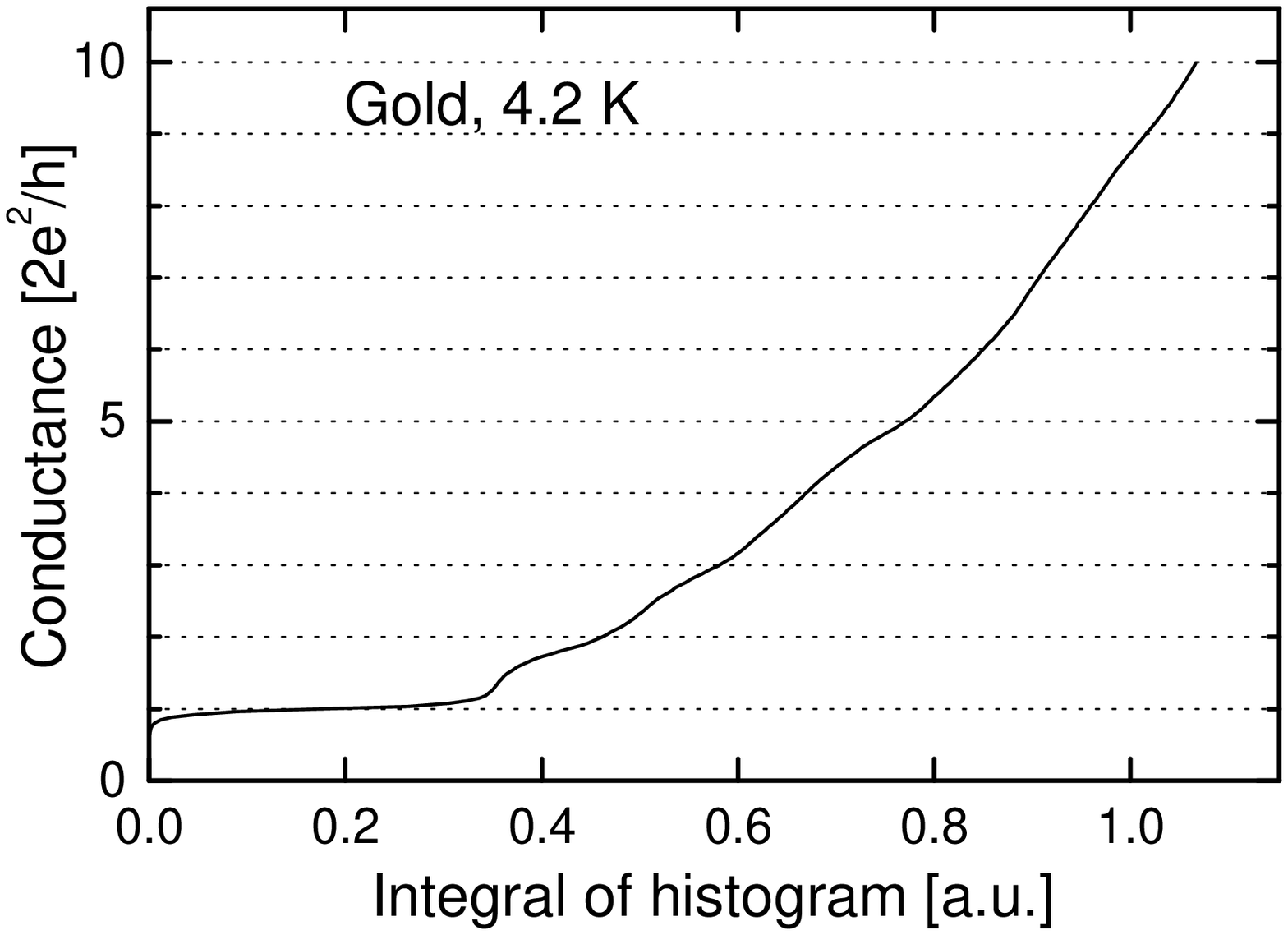}
\caption{Curve obtained by integrating a gold histogram similar to Fig.~\protect\ref{fig:gold} (from A.I. Yanson \protect\cite{yanson98b})}
\label{fig:integrated}
\end{figure}

When we assume that the contact breaking process produces any effective contact diameter with equal probability, then the histograms represent a derivative of conductance with respect to the effective diameter of the contact. It is instructive to calculate the integral of the histogram, as was first done by Gai et al.\   \cite{gai}. Fig.~\ref{fig:integrated} shows such a curve, obtained from a gold histogram similar to the one shown in Fig.~\ref{fig:gold}. This curve is to be compared to conductance traces obtained for 2D semiconductor devices \cite{wees,wharam}, for which the width of the contact can directly and continuously be adjusted by the gate electrostatic potential. Compared to the latter, the conductance steps in Fig.~\ref{fig:integrated} are poorly defined, with the exception of the first conductance quantum. Moreover, we will show that the first quantum feature results from the fact that our assumption mentioned above is not valid. The effective diameters produced during contact breaking are strongly influenced by the possible atomic configurations. As we will argue below, the step at 1~$G_0$, corresponding to the strong peak in the histogram for gold, results from the formation of a chain of gold atoms\index{Atomic chain} \index{Chain of atoms} during the last stages of contact breaking. Disregarding the first step in Fig.~\ref{fig:integrated}, we find that the conductance is not strictly quantised, as the probability of finding a contact with a conductance of, e.g., 2~$G_0$ is only twice that of finding 1.5~$G_0$. However, the conductance is still determined by the quantum states, as described in Sect.~\ref{sec:theory} and is carried by a limited number of modes. We will show below how the quantum nature for monovalent metals is revealed by a tendency for the modes to open one-by-one as the contact becomes larger.

\section{The Character of the Conductance Modes Through a Single Atom}
\index{Conductance mode}
\index{Conductance channel}
Instead of discussing the average properties of many contacts, we now concentrate on the simplest configuration, a single atom bridging the electrodes at either side, and consider the question of what the conductance for this atomic configuration will be. Ideally, we would like to know the number of modes contributing to the conductance and the transmission probability, $T_n$ for each of these. From a measurement of the conductance alone we cannot obtain this information, since the conductance gives only the sum of $T_n$. Scheer et al.\   \cite{scheer97} have introduced a method which allows us to obtain this information from experiment. The method exploits the non-linearities in the current--voltage characteristic for contacts in the superconducting state. Other techniques, which give more limited information on the contribution of the various modes, will be discussed at the end of this section.

\subsection{Subgap structure in superconducting contacts}
\begin{figure}[!t]
\includegraphics[height=.9\textwidth,angle=90]{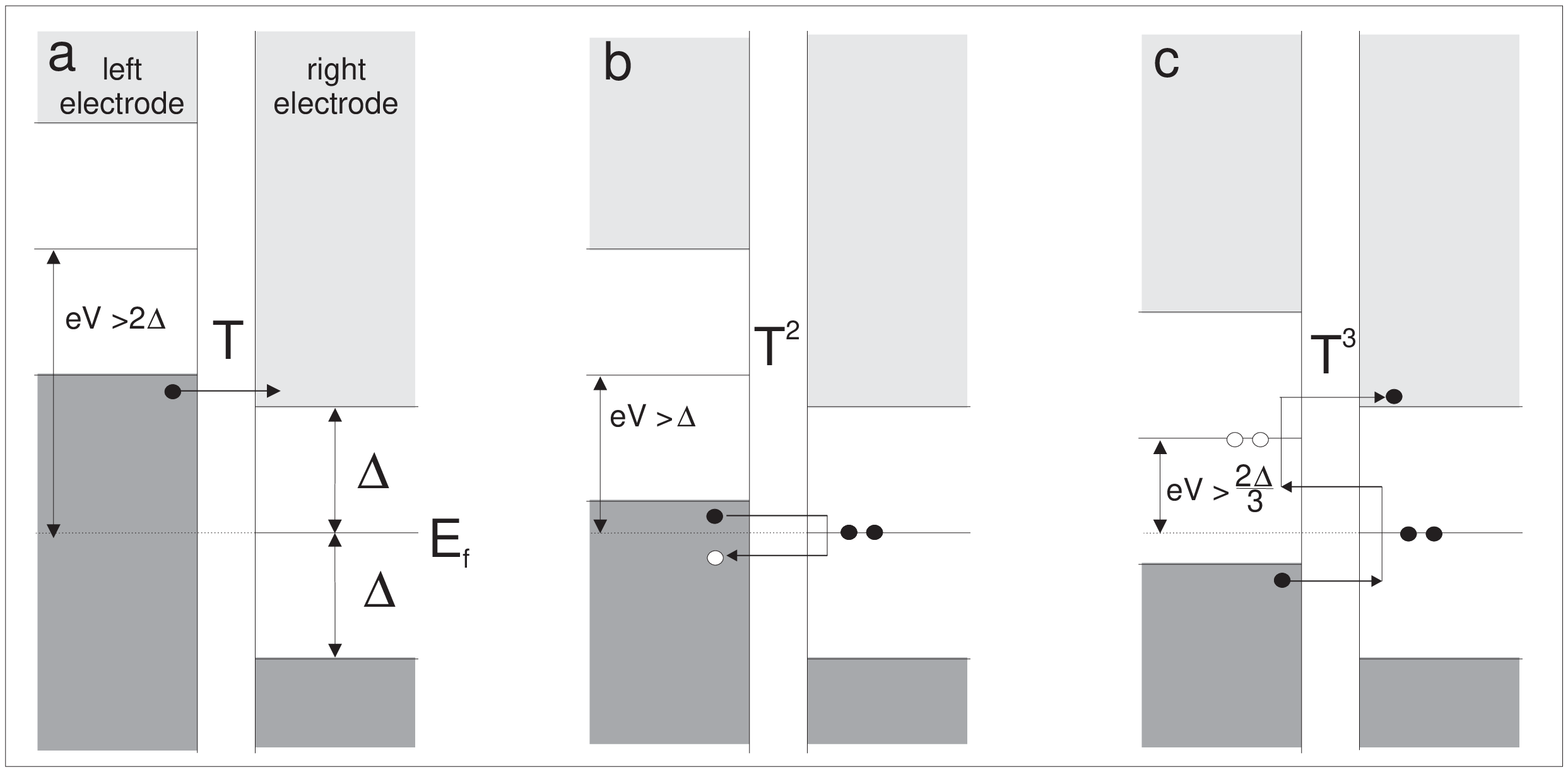}
\caption{Illustration of the Multiple Particle Tunnelling\index{multiple particle tunnelling} (MPT), or Multiple Andreev Reflection\index{multiple Andreev reflection} (MAR) processes. In each of the three diagrams the available quasi-particle states as a function of energy ({\it vertical axis}) at both sides of the tunnel barrier are given in the semiconductor representation. {\it Dark grey} are the occupied states, {\it light grey} the unoccupied states and the {\it line} in the middle of the gap represents the Cooper pair energy, which is separated by an energy $\Delta$ from the occupied and the unoccupied states. Applying an external electrical potential $V$ across the junction shifts the states on the left side of the junction up by an energy $eV$ with respect to those in the right electrode. The ordinary superconducting tunnelling process is given in ({\bf a}), which shows that a voltage $V>2\Delta/e$ is required for single quasi-particles to cross the junction. The probability for tunnelling of a particle, determined by the transparency of the barrier, is $T$. This gives rise to the familiar jump in the current at $eV=2\Delta$ in the current--voltage characteristic of a superconducting tunnel junction. When we consider higher order processes, the next order is represented in ({\bf b}). This can be described as two quasi-particles crossing simultaneously to form a Cooper pair in the right electrode (MPT). Alternatively, the process can be regarded as an electron-like quasiparticle falling onto the barrier, which is reflected as a hole-like quasiparticle, forming a Cooper pair on the right (MAR). The two descriptions are equivalent, and give rise to a current step in the 
current--voltage characteristic of the junction at $eV=2\Delta/2$. The probability for the process is $T^2$, since it requires the crossing of two particles. ({\bf c}) shows the third order process, which involves breaking up a Cooper pair on the left, combining it with a quasiparticle, to form a Cooper pair and a quasiparticle on the right.  It is allowed for $eV>2\Delta/3$ and has a probability $T^3$ }
\label{fig:MPT}
\end{figure}

\index{Subgap structure}
\index{Superconducting contact}
The principle of the method introduced by Scheer et al.\  can be illustrated by considering first a contact having a single mode with low transmission probability, $T\ll 1$. For $T\ll 1$ we have essentially a tunnel junction, and the current--voltage characteristic\index{Current-voltage characteristic}
 for a superconducting tunnel junction\index{Superconducting tunnel junction} is known to directly reflect the gap, $\Delta$, in the density of states for the superconductor \cite{wolf}. As illustrated in Fig.~\ref{fig:MPT}a no current flows until the applied voltage exceeds $2\Delta/e$ (where the factor 2 results from the fact that we have identical superconductors on both sides of the junction), after which the current jumps to approximately the normal-state resistance line. For $eV>2\Delta$ single quasi-particles can be transferred from the occupied states at $E_{\rm F} - \Delta$ on the low voltage side of the junction to empty states at $E_{\rm F} + \Delta$ at the other side. For $eV<2\Delta$ this process is blocked, since there are no states available in the gap.\index{Superconducting energy gap}

However, when we consider higher order tunnel processes a small current can still be obtained. Figure~\ref{fig:MPT}(b) illustrates a process, which is allowed for $eV > \Delta$ and consists of the simultaneous tunnelling of {\it two} quasiparticles from the low  bias side to form a {\it Cooperpair} on the other side of the junction. The onset of this process causes a step in the current at half the gap value, $V= 2\Delta/2e$. The height of the current step is smaller than the step at $2\Delta/e$ by a factor $T$, since the probability for two particles to tunnel is $T^2$. In general, one can construct similar processes of order $n$, involving the simultaneous transfer of $n$ particles, which give rise to a current onset at $eV=2\Delta/n$ with a step height proportional to $T^n$. An example for $n=3$ is illustrated in Fig.~\ref{fig:MPT}c. This mechanism is known as multiple particle tunnelling\index{multiple particle tunnelling} and was first described by Schrieffer and Wilkins \cite{MPT}. It is now understood that this is the weak coupling limit of a mechanism which is referred to as multiple Andreev reflection
\index{multiple Andreev reflection}  \cite{KBT,arnold,averin,cuevas96,bratus95,bratus97}. The theory could only be tested recently, since it requires the fabrication of a tunnel junction having a single tunnelling mode with a well-defined tunnelling probability $T$. For atomic size niobium\index{Nb} tunnel junctions the theory was shown to give a very good agreement \cite{vdpost94,vdpost97}, describing up to three current steps, including the curvature and the slopes, while the only adjustable parameter is the tunnel probability, which follows directly from the normal state resistance.

\begin{figure}[!t]
\includegraphics[height=.8\textwidth,angle=270]{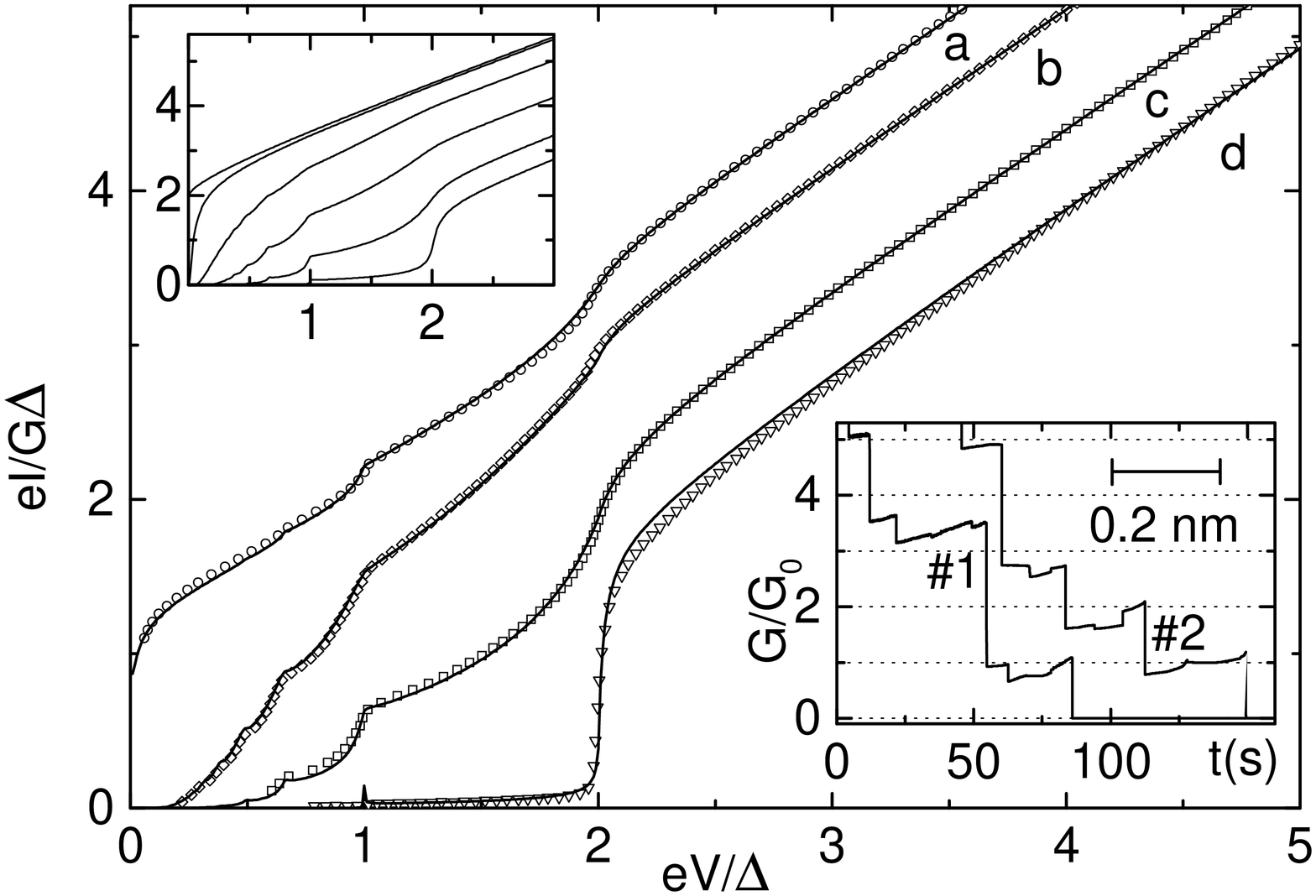}
\caption{Current--voltage characteristics for four atom-size contacts of aluminium\index{Al} using a lithographically fabricated mechanically controllable break junction at 30~mK ({\it symbols}). The {\it right inset} shows the typical variation of the conductance, or total transmission $T=G/G_0$, as a function of the displacement of the electrodes, while pulling, and is similar to the curves for aluminium shown in Fig.~\protect\ref{fig:aluminium}. The bar indicates the approximate length scale. The data in the main panel have been recorded by stopping the elongation at the last stages of the contact ({\bf a}--{\bf c}) or just after the jump to the tunnelling regime ({\bf d}) and then measuring the current while slowly sweeping the bias voltage. The current and voltage are plotted in reduced units, $eI/G\Delta$ and $eV/\Delta$, where $G$ is the normal state conductance for each contact and $\Delta$ is the measured superconducting gap, $\Delta/e=(182.5\pm2.0)\umu$V. The {\it left inset} shows the current--voltage characteristics obtained from first-principles theory for a single channel junction \protect\cite{averin,cuevas96,bratus95,bratus97} with different values for the transmission probability $T$ (from bottom to top: $T$=0.1, 0.4, 0.7, 0.9, 0.99, 1). The {\it full curves} in the main panel have been obtained by adding several theoretical curves and optimising the set of $T$ values. The curves are obtained with: ({\bf a}) three channels, $T_1$=0.997, $T_2$=0.46, $T_3$=0.29 with a total transmission $\sum T_n =$1.747, ({\bf b}) two channels, $T_1$=0.74, $T_2$=0.11, with a total transmission $\sum T_n =$0.85, ({\bf c}) three channels, $T_1$=0.46, $T_2$=0.35, $T_3$=0.07 with a total transmission $\sum T_n =$0.88. ({\bf d}) In the tunnelling range a single channel is sufficient, here $\sum T_n=T_1$=0.025.
(From Scheer et al.\   \protect\cite{scheer97})}
\label{fig:AlSubGap}
\end{figure}

Since the theory has now been developed to all orders in $T$ \cite{arnold,averin,cuevas96,bratus95,bratus97}, Scheer et al.\  \cite{scheer97} realised that this mechanism offers the possibility of extracting the transmission probabilities\index{Transmission probability} for contacts with a finite number of channels contributing to the current, and is ideally suited to analysing atomic size contacts. Roughly speaking, the current steps at $eV=2\Delta/n$ are proportional to $\sum T_m^n$, with $m$ the channel index, and when we can resolve sufficient details in the current--voltage characteristics, we can fit many independent sums of powers of $T_m$'s. When the $T_m$'s are not small compared to 1, all processes to all orders need to be included for a description of the experimental curves. In practice, the full expression for the current--voltage characteristic for a single channel from theory \cite{averin,cuevas96,bratus95,bratus97} is numerically evaluated for a given transmission probability $T_m$ (Fig.~\ref{fig:AlSubGap}, inset), and a number of such curves are added independently, where the $T_m$'s are used as fitting parameters. Scheer et al.\  tested their approach first for aluminium\index{Al} contacts \cite{scheer97}. As shown in Fig.~\ref{fig:AlSubGap}, all 
Current--voltage curves for small contacts can be very well described by the theory. However, the most important finding was that at the last ``plateau'' in the conductance, just before the breaking of the contact, typically three channels with different $T$'s are required for a good description, while the total conductance for such contacts is of order 1~$G_0$, and would in principle require only a single conductance channel. Contacts at the verge of breaking are expected to consist of a single atom, and this atom would then admit three conductance channels, but each of the three would only be partially open, adding up to a conductance close to 1~$G_0$. This very much contradicts a simple picture of quantised conductance in atomic size contacts, and poses the question as to what determines the number of channels through a single atom.

\subsection{Valence-Orbital-Based Description of the Conductance Modes}
\begin{figure}[!t]
\includegraphics[width=.7\textwidth]{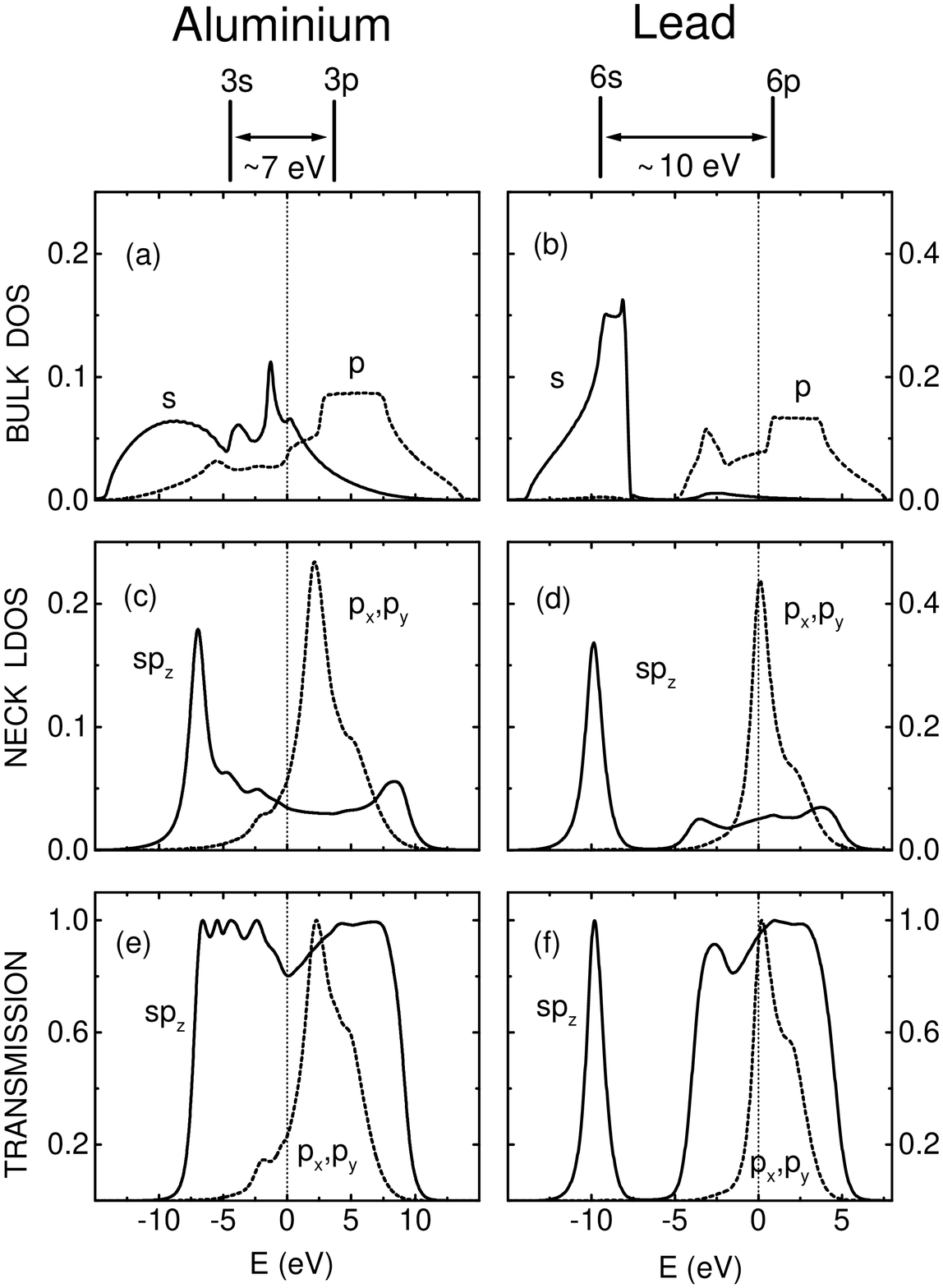}
\caption{Localized orbital model for electrical conduction through 
one-atom contacts. The atomic valence levels ({\it vertical bars} above the figures) develop into bulk conduction bands
({\bf a}) and ({\bf b}) for Al\index{Al} and Pb, respectively.\index{Pb} The {\it panels} ({\bf c}) and ({\bf d}) depict schematically the local density of states (LDOS) in eV$^{-1}$ at the central atom of the model geometry for a single atom contact. The global energy dependence of the transmission coefficients $T_n$ is shown in ({\bf e}) and ({\bf f}). The {\it dotted lines} indicate the position of the Fermi level. The $sp_z$ mode is the best transmitted for both materials. The $p_x$ and $p_y$ modes are degenerate due to the symmetry of the model geometry. (From \protect\cite{cuevas98,scheer98})}
\label{fig:cuevas}
\end{figure}

\index{Valence orbital}
\index{Conductance mode}
\index{Conductance channel}
Cuevas, Levy Yeyati and Mart\' \i n-Rodero \cite{cuevas98} constructed a model to explain these results, using a tight binding calculation,\index{Tight binding calculation} for a geometry of two atomic pyramids touching at the apex through a single atom.
They argue that it is very important to make the tight binding calculation self-consistent, by which they mean that local charge neutrality is maintained at each atomic site, by iteration and adjustment of the site energy for each individual atom in the model configuration. Figure~\ref{fig:cuevas} shows the results of their calculations of the density of states and the transmission probability for the various channels as a function of energy. They find that the conductance channels can be described in terms of the atomic valence orbitals.\index{Valence orbital}
 Aluminium has a configuration [Ne]$3s^2 3p^1$, and a total of four orbitals would be available for current transport: one $s$ orbital and three $p$ orbitals, $p_x, p_y$ and $p_z$. They identify in their calculation three contributions, one which originates from a combination of $s$ and $p_z$ orbitals (where the $z$ coordinate is taken in the current direction), and two smaller identical contributions labelled $p_x$ and $p_y$. The degeneracy of these two channels is due to the symmetry of the problem, and can be lifted by changing the local environment for the central atom. The fourth possible channel, an antisymmetric combination of $s$ and $p_z$, is found to have a negligible transmission probability. Thus, their calculation confirms the experimental observation by Scheer et al.\  that three channels contribute to the conductance for a single aluminium atom. It was also found that the total conductance for the three channels is of order 1~$G_0$. The results are very robust against changes in the atomic configuration; only the total conductance varies somewhat between different choices for the atomic geometry. The qualitative features of the model agree with ab initio calculations for single atom contacts and a simplified structure for the environment \cite{lang87,lang95,lang97,wan97}.

\begin{figure}[!t]
\includegraphics[width=.8\textwidth]{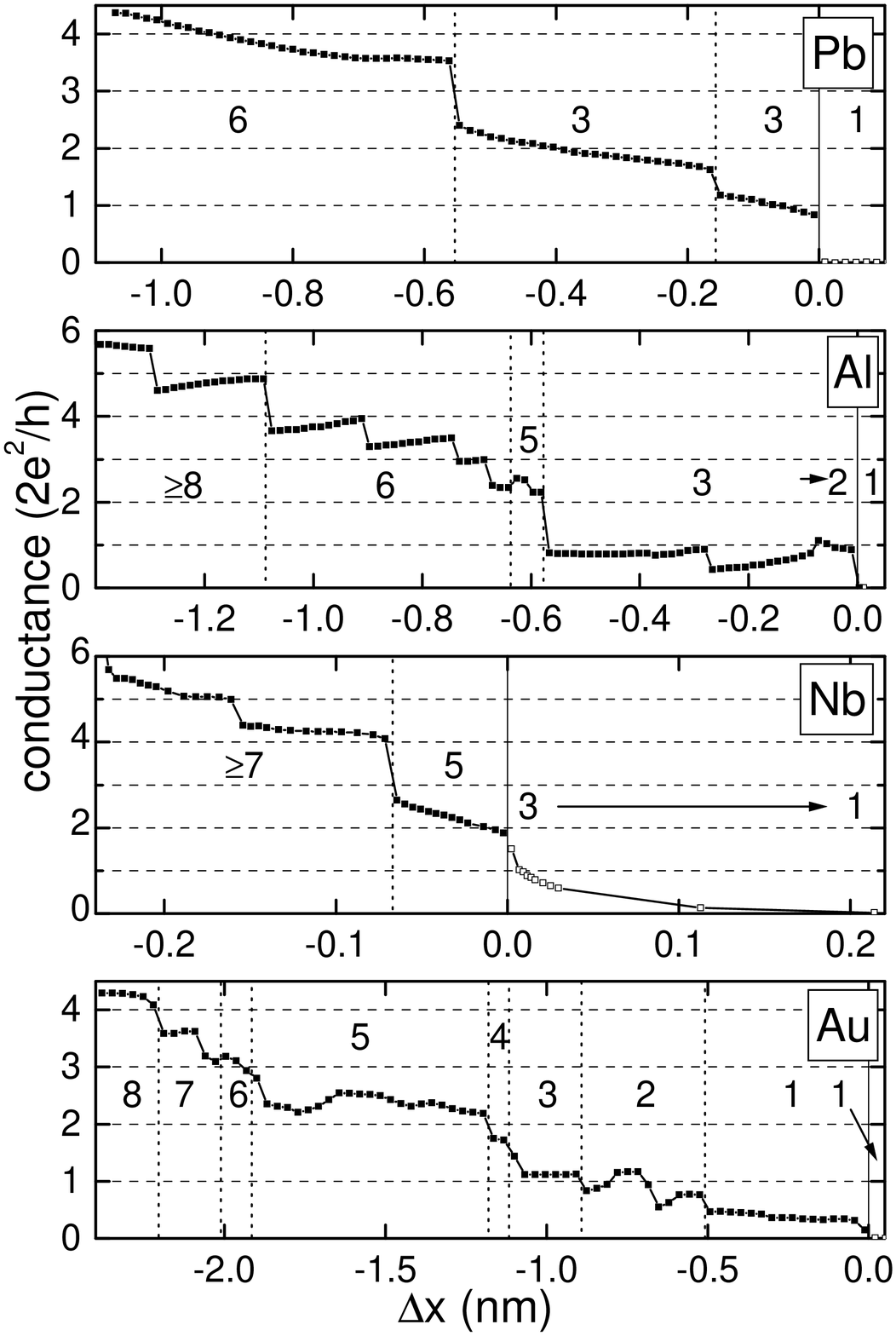}
\caption{Conductance curves measured as a function of contact elongation for Al,\index{Al} Pb,\index{Pb} Nb\index{Nb} and Au.\index{Au} The number of channels contributing to the conductance was determined at each point in the curves by recording the current--voltage relation and fitting the curves with the theory for superconducting subgap structure. The {\it numbers along the curves} in the figure indicate the number of channels obtained in this way. The number is constant over a plateau, and usually jumps to a smaller value at the steps in the conductance. (From \cite{scheer98})}
\label{fig:scheer98}
\end{figure}

The analysis, both theoretical and experimental, was extended to other metals \cite{cuevas98,scheer98}, by which it was shown that the number of conductance channels for an atom of a given metallic element depends on the number of valence orbitals. Figure~\ref{fig:scheer98} shows conductance curves, similar to those in Fig.~\ref{fig:gold}, for Pb,\index{Pb} Al,\index{Al} Nb\index{Nb} and Au,\index{Au} where at each point in the figure current--voltage curves as in Fig.~\ref{fig:AlSubGap} were recorded and fitted in order to determine the number of channels involved. This number of channels is indicated along the curves in Fig.~\ref{fig:scheer98}. The number is constant over a plateau in the conductance, where the transmission probability for every mode changes gradually. At the steps in the conductance the number of channels involved is usually found to jump to a smaller number. In tunnelling range, when the contact is broken and the distance is larger than 0.2~nm, the current--voltage characteristics\index{Current-voltage characteristic} can in all cases be described by a single channel, with a transmission probability which is given by the tunnelling resistance. The number of channels found for the smallest contacts, just before the jump to tunnelling is 1 for Au, 3 for Al and Pb, and 5 for Nb. Note that gold is not a superconductor, and a special device was fabricated which allowed the use of proximity induced superconductivity\index{Proximity effect} \cite{scheer98}. The device is a nanofabricated version of a break junction, having a thick superconducting aluminium layer forming a bridge with a gap of about 100~nm. This small gap was closed by a thin gold film in intimate contact with the aluminium. Superconducting properties were thereby induced in the gold film, and by breaking the gold film and adjusting an atomic size contact, the same subgap analysis could be performed. Both the Al and Au junctions were measured at temperatures of 100~mK, far below the superconducting transition temperatures. Pb and Nb were measured at 1.5~K.

The number of channels and the total conductance at the last plateau before breaking found in the experiment agree very well with the theory, for which 
the results can be summarised in the following way. Single atom contacts for monovalent metals, including the noble metals Cu, Ag and Au, and the alkali metals, have a single valence orbital available for current transport, giving rise to a single channel with a transmission probability close to unity. The total conductance for such contacts is thus expected to be close to 1~$G_0$. The experiment for gold shows indeed a single channel, but the total conductance is a factor 2--3 smaller than predicted. This was tentatively attributed to the strong scattering in the nano-fabricated device, and other techniques, which will be discussed below, have found more close agreement with the predicted total conductance of 1~$G_0$.
For $s$-$p$ metals, including Al and Pb, three channels give a noticeable contribution to the current. The fourth channel, i.e., the antisymmetric combination of $s$ and $p_z$ is always nearly closed. The total transmission depends on the number of valence {\it electrons}. Pb has 4 valence electrons while aluminium has only three, and this has roughly the effect that the Fermi energy in Fig.~\ref{fig:cuevas} is shifted up to the peak in the $p_x$,$p_y$ channel for Pb. The total conductance for Pb is found to be nearly 3~$G_0$ in the calculations, while the conductance at the last plateau for lead does indeed start close to three quantum units, dropping to lower values upon further stretching of the contact. Niobium is a $d$-metal with a configuration [Kr]$4d^{4}5s^1$ having 6 valence orbitals: 1~$s$ and 5~$d$. The theory again predicts one combination with a negligible contribution and that the remaining five channels should add up to a total conductance of about 2.8~$G_0$, again in good agreement with the experiment. The overall agreement is good, but some variation is observed between contacts and the information is obtained on a limited number of contacts. Below techniques are discussed which give less detailed information, but directly show the average properties of the contacts, and confirm the picture presented here. 

\subsection{Further Experimental Techniques}

Three other methods have recently been introduced in order to investigate the number of channels contributing to the conductance of atomic size contacts: the measurement of shot noise, conductance fluctuations and thermopower.

\subsubsection*{Shot noise}\index{Shot noise} is the result of the discrete character of the current due to the passage of individual electrons. It was originally found in vacuum diodes, and first discussed by Schottky in 1918 \cite{schottky}. The passage of individual electrons can be regarded as a delta function of the current with time. The total current is the sum of a random distribution of such delta functions, giving a time averaged current $I$, and a frequency spectrum of fluctuations which is white (up to very high frequencies), with a noise power equal to $2eI$. This shot noise can be observed, e.g., in tunnel junctions. 

For a perfect ballistic point contact, in the absence of back-scattering, i.e., all channel transmission probabilities are either 1 or 0, the shot noise is expected to vanish
\cite{lesovik,butt90,been91,martin92,butt92,scherbakov}. This can be understood from the wave nature of the electrons, for which the wave function extends from the left bank to the right bank of the contact without interruption. When the state on the left is occupied for an incoming electron, it is occupied on the right as well and there are no fluctuations in this occupation number. In other words, the incoming electron is not given the choice of being transmitted or not, it is always transmitted when it enters an open mode. In order to have noise, the electron must be given the choice of being reflected at the contact. This will be the case when the transmission probability is smaller than 1 and larger than 0. In single-channel quantum point contacts, shot noise is predicted to be suppressed by a factor proportional to $T(1-T)$, where $T$ is the transmission probability\index{Transmission probability} of the conductance channel
\cite{lesovik,butt90,been91,martin92,butt92}. This quantum suppression has recently been observed in point contact devices in a 2-dimensional electron gas \cite{reznikov,kumar}. For a general multichannel contact the shot noise power is predicted to be
\begin{equation}
P_I = 2eV G_0 \sum_n T_n (1-T_n)\; .  \label{eq:sn} 
\end{equation}
Since this depends on the sum over the second power of the transmission coefficients, this quantity is independent of the conductance, $G=G_0\sum T_n$, and simultaneous measurement of these two quantities should give information about the channel distribution.

\begin{figure}[!t]
\includegraphics[width=.8\textwidth]{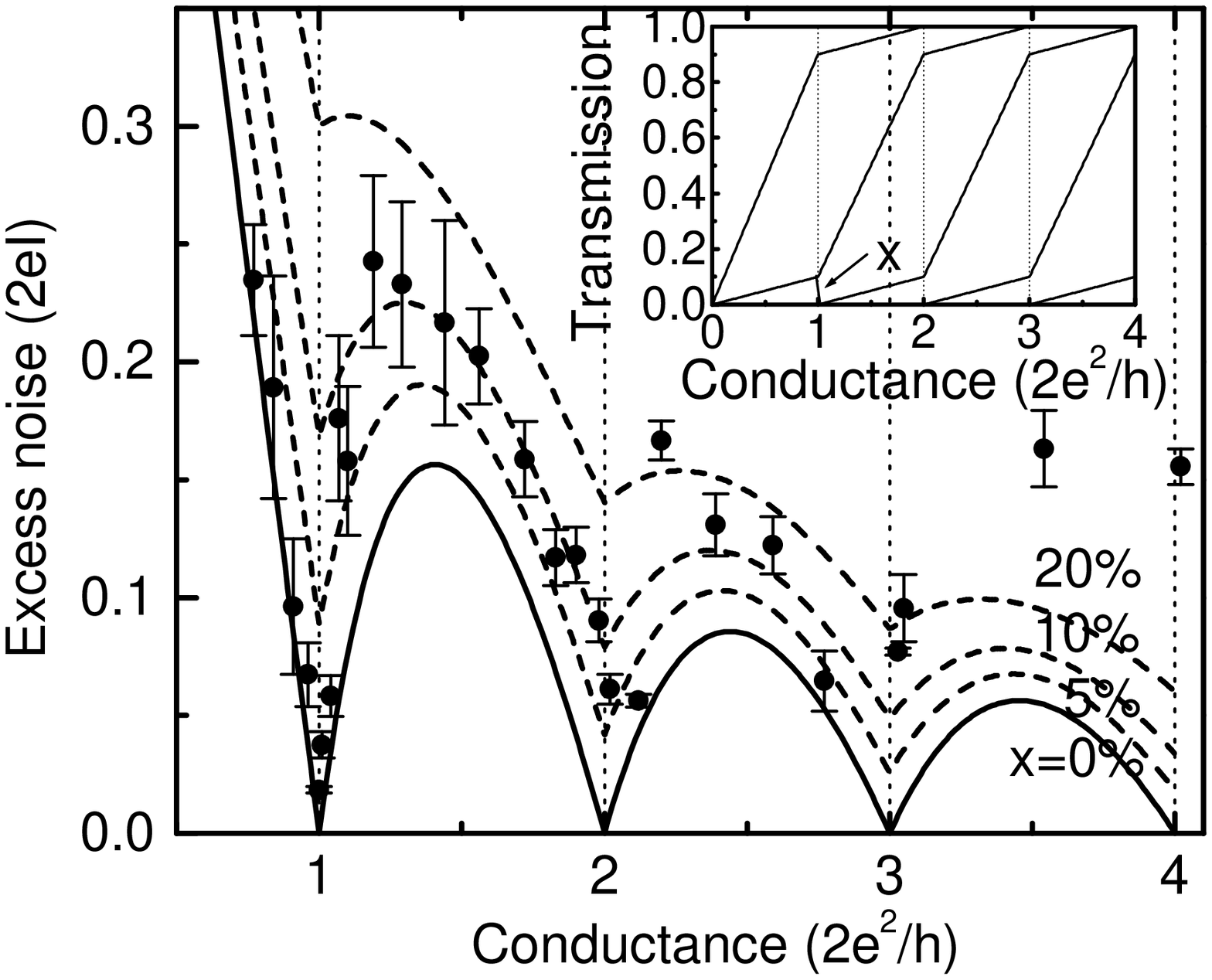}
\caption{Measured shot noise values for 27 gold\index{Au} contacts at 4.2\,K with a bias current of 0.9\,$\umu$A. Comparison is made with calculations in the case of one single partially transmitted mode ({\it full curve}) and for various amounts of contributions of other modes according to the model described in the inset ({\it dashed curves}). In the limit of zero conductance, these curves all converge to full shot noise, i.e.,  2.9$\times 10^{-25}\,$A$^2$/Hz.
(From \cite{brom99})}
\label{fig:shotnoise}
\end{figure}

When measuring shot noise on atomic size point contacts \cite{brom99} it is necessary to work at low temperatures in order to reduce thermal noise, and to shield the contact carefully from external mechanical and acoustic vibrations. Using two sets of preamplifiers in parallel and measuring the cross-correlation of the noise for the two signals eliminates the noise of the preamplifier. Experimental results for shot noise in gold\index{Au} point contacts for a number of conductance values are shown in Fig.~\ref{fig:shotnoise}. From the two measured parameters, $G$ and $P_I$, one can determine at most two independent transmission probabilities. Instead, the results in Fig.~\ref{fig:shotnoise} are compared to models which assume a certain evolution of the values for $T_n$ as a function of the total conductance, $G$. The full curve shows the simplest model, where all $T_n$ are either 0 or 1, except for a single partially open channel, so that the total conductance is given by $G/G_0=N-1 + T_N$. The figure shows that the experimental data closely follow this behaviour, in particular at low conductance. In order to estimate the contribution of additional partially open channels the broken curves show the expected behaviour for an evolution of the channel transmissions as illustrated in the inset. It shows that the deviation from a model with the channels opening one-by-one is only 10\% between 1 and 2~$G_0$, and about 20\% between 2 and 3~$G_0$. The scatter in the data points is larger than the experimental error because each point was measured on a different contact, which can have a very different set of transmission values, and a purely systematic behaviour is not expected. Although we cannot determine each individual transmission value, we obtain information from the property that the noise increases the more channels are partially transmitted. The minimum possible noise for a given conductance value is represented by the full curve in Fig.~\ref{fig:shotnoise}, corresponding to a single partially open channel. The broken curves give an impression of the other partially open channels.

There is a very strong suppression, down to 2\% of the full shot noise value, for $G=1~G_0$. It needs to be stressed that this holds for gold contacts. There is a fundamental distinction between this monovalent metal and the multivalent metal aluminium,\index{Al} which shows no systematic suppression of the shot noise at multiples of the conductance quantum, and the intensities lie between 0.3 and 0.6 times $2eI$ for $G$ close to $G_0$ \cite{brom99}.

\subsubsection*{Conductance Fluctuations.}
\begin{figure}[!t]
\includegraphics[width=.8\textwidth]{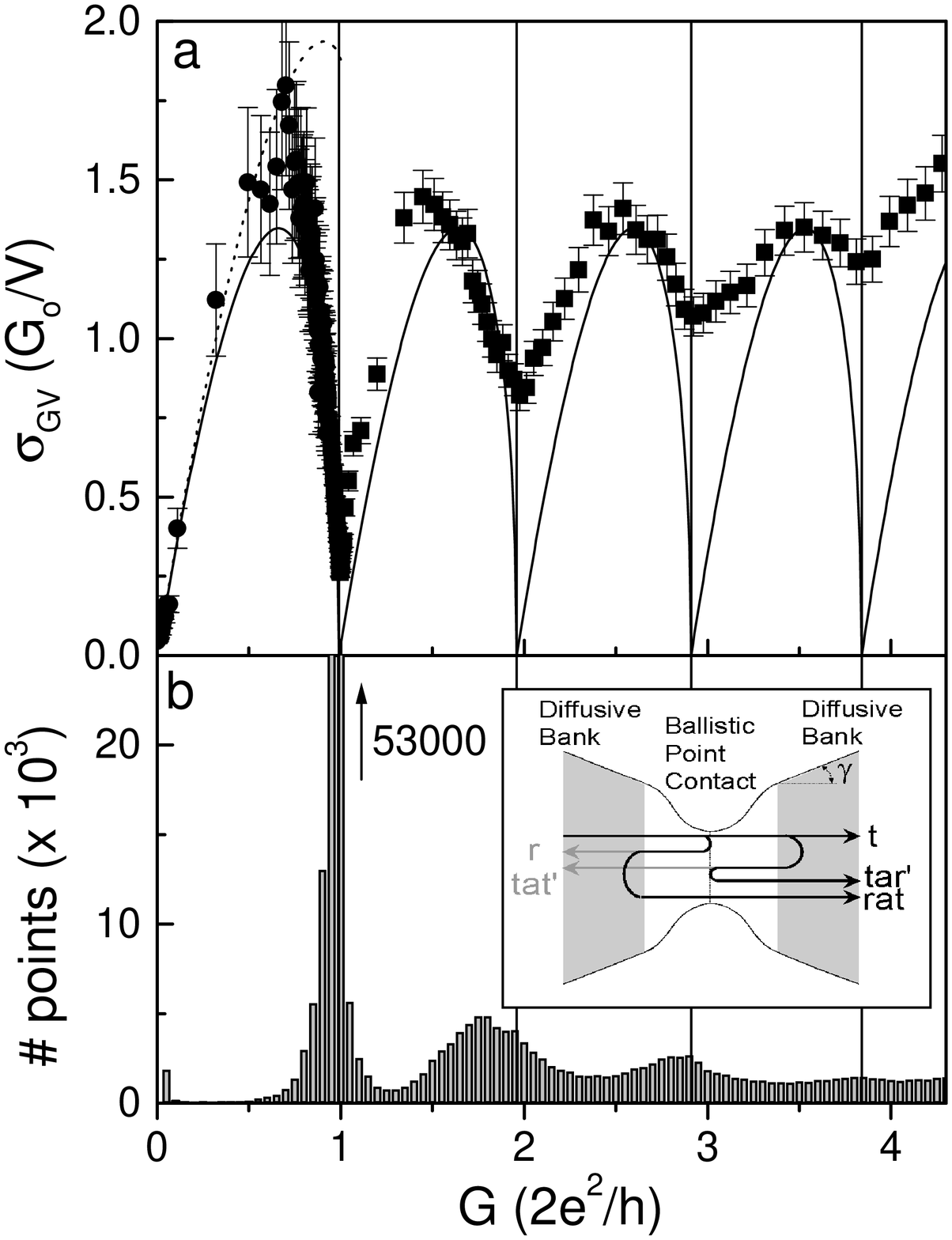}
\caption{({\bf a}) Standard deviation of the voltage dependence of the conductance versus conductance for 3500 curves for gold.\index{Au} All data points in the set were sorted as a function of the conductance after which the rms value of ${\rm d}G/{\rm d}V$ was calculated from a fixed number of successive points. The {\it circles} are the averages for 300 points, and the {\it squares} for 2500 points. The {\it solid} and {\it dashed curves} depict the calculated behaviour for a single partially-open channel and a random 
distribution over two channels respectively. The {\it vertical grey lines} are the corrected integer conductance values (see text). ({\bf b}) Conductance histogram obtained from the same data set. The peak in the conductance histogram at $G_{0}$ extends to 53000 on the {\it y}-scale. ({\it Inset}) Schematic diagram of the configuration used in the analysis. The {\it dark lines} with arrows show the paths, which contribute to the conductance fluctuations in lowest order. (From \cite{ludoph99})}
\label{fig:2deriv}
\end{figure}

\index{Conductance fluctuations}
Interference between electron trajectories scattering on defects near atomic size metallic contacts, gives rise to dominant contributions to the second derivative of the current with respect to bias voltage $V$, i.e., in ${\rm d}G/{\rm d}V$. This effect has the same origin as the well-known Universal Conductance Fluctuations (UCF) in diffusive mesoscopic conductors \cite{spivak}. It was studied for point contacts an order of magnitude larger than the atomic size by Holweg et al.\  \cite{holweg91,holweg93} and by Ralph et al.\  \cite{ralph93}, and the theory was developed by Kozub et al.\  \cite{kozub94}.  In experiments on gold contacts \cite{ludoph99} it was found that this voltage dependence is suppressed near multiples of the quantum value of conductance, $n(2e^2/h)$. By applying a constant modulation voltage at frequency $\omega$ and measuring the current with lock-in amplifiers simultaneously at $\omega$ and the second harmonic $2\omega$ the conductance and its derivative can be obtained during conductance scans as in Fig.~\ref{fig:gold}. From the combined data sets of many such curves one can construct a conductance histogram together with the average properties of ${\rm d}G/{\rm d}V$. It was found that ${\rm d}G/{\rm d}V$ for a given conductance value has a bell-shaped distribution centred around zero and that the width of this distribution has a systematic variation with conductance. Figure~\ref{fig:2deriv} shows the standard deviation of the derivative of the conductance with bias voltage $\sigma_{GV}=\langle ({\rm d}G/{\rm d}V)^2 \rangle$, obtained from 3500 curves for gold, where the conductance and the derivative of the conductance were measured simultaneously as a function of contact elongation. The conductance histogram\index{Conductance histogram} for the same set of data is shown in the lower panel.

The data for $\sigma_{GV}$ display pronounced minima for $G$ near multiples of $G_0$. A similar effect can be recognised in the numerical calculations of Maslov et al.\   \cite{maslov}. The explanation for this quantum suppression of the conductance fluctuations as presented in \cite{ludoph99} is illustrated in the inset of Fig.~\ref{fig:2deriv}. The contact is modelled by a ballistic central part, which can be described by a set of transmission values for the conductance modes, sandwiched between diffusive banks, where electrons are scattered by defects characterised by an elastic scattering length $l_{\rm e}$. An electron wave of a given mode falling onto the contact is transmitted with probability amplitude $t$ and part of this wave is reflected back to the contact by the diffusive medium, into the same mode, with probability amplitude $a_n\ll 1$. This back-scattered wave is then {\it reflected} again at the contact with probability amplitude $r_n$, where $T_n = |t_n|^2= 1-|r_n|^2$. The latter wave interferes with the original transmitted wave. This interference depends on the phase difference between the two waves, and this phase difference depends on the phase accumulated by the wave during the passage through the diffusive medium. The probability amplitude $a_n$ is a sum over all trajectories of scattering, and the phase for such a trajectory of total length $L$ is simply $kL$, where $k$ is the wave vector of the electron. The wave vector can be influenced by increasing the voltage over the contact, thus launching the electrons into the other electrode with a higher speed. The interference of the waves changes as we change the bias voltage, and therefore the total transmission probability, or the conductance, changes as a function of $V$. This describes the dominant contributions to the conductance fluctuations, and from this description it is clear that the fluctuations are expected to vanish either when $t_n=0$, or when $r_n=0$. 

Elaborating this model Ludoph et al.\  obtained the following analytical expression for $\sigma_{GV}$,
\begin{equation} 
\sigma_{GV} = \frac {2.71\ e\ G_0}{\hbar k_{\rm F} v_{\rm F} \sqrt{1-\cos\gamma}} 
\Bigl(\frac {\hbar/\tau_{\rm e}}{eV_{\rm m}}\Bigr)^{3/4}
\sqrt{\sum_n T_n^{2}(1-T_n)}\; , \label{eq:sigmaGV}
\end{equation}
where $k_{\rm F}$ and $v_{\rm F}$ are the Fermi wave vector and Fermi velocity, respectively, $\tau_{\rm e}=l_{\rm e}/v_{\rm F}$ is the scattering time. The shape of the contact is taken into account in the form of the opening angle $\gamma$ (see the inset in Fig.~\ref{fig:2deriv}), and $V_{\rm m}$ is the applied voltage modulation amplitude. The full curves in Fig.~\ref{fig:2deriv}a are obtained from (\ref{eq:sigmaGV}), assuming a single partially-open channel at any point, i.e., assuming that channels open one-by-one as the conductance increases. In agreement with the results discussed above, the conductance for the smallest gold contacts is very well described by this simple approximation. The amplitude of the curves is adjusted to fit the data, from which a value for the mean free path\index{Mean free path} is obtained, $l_{\rm e}=5\pm 1$~nm. Similar experiments \cite{ludoph99,ludoph99b}  for copper\index{Cu} and silver\index{Ag} and for sodium\index{Na} also show the quantum suppression of  conductance fluctuations observed here for gold, while for aluminium\index{Al} or niobium\index{Nb} it is not observed. 

The back-scattering, which produces the conductance fluctuations, has a second observable effect, namely on the average conductance, as opposed to the fluctuating part. This effect is seen as a shift to lower conductance of the minima in $\sigma_{GV}$, and can be described to lowest order by an effective resistance in series with the contact. This shift has been taken into account in the calculation of the curves in Fig.~\ref{fig:2deriv}a. Also in the interpretation of conductance histograms, a phenomenological series resistance is often taken into account in order to describe the shift of the peaks to lower values \cite{krans95,hansen,costa97a,costa97b}. The estimate of the mean free path obtained from this series resistance is indeed consistent with the value obtained by fitting expression (\ref{eq:sigmaGV}) to the data.

\subsubsection*{Thermopower.}\index{Thermopower} Further information on the quantum transport properties of atomic size contacts, again for gold,\index{Au} has been obtained from thermopower experiments using a modified MCB technique \cite{ludoph99c}. By applying a constant temperature difference over the contacts, the thermally induced potential could be measured simultaneously with the conductance. Large thermopower values were obtained, which jump to new values simultaneously with the jumps in the conductance. The values are randomly distributed around zero with a roughly bell-shaped distribution, in contrast to what was expected from elementary free electron gas models \cite{bogachek96}. The thermopower was shown to result from the same mechanism as the conductance fluctuations, and the theory was adapted to describe the results. The experimental results were found to follow the law obtained from this defect-scattering model, and quantum suppression of the thermopower was indeed observed. By fitting the curves, a value for the mean free path was obtained which was in close agreement with the value from the conductance fluctuations.\index{Conductance fluctuations} Since the two experimental techniques are very different, and the typical energy scales for the effects are at least an order of magnitude apart (this scale is set by the modulation voltage amplitude, 20~mV, in one case and the temperature, $\sim$10~K, equivalent to $\simeq 1$~mV, in the other), these results give strong support for the description and interpretation presented above.

\subsection{Implications for Conductance Curves and Histograms}
Let us first focus on results for the monovalent metal, gold, which has been most studied. The results obtained from superconducting subgap structure,\index{Subgap structure} \index{Superconducting contact} shot noise\index{Shot noise} measurements, conductance fluctuations\index{Conductance fluctuations} and thermopower\index{Thermopower} are in agreement and show that the conductance for the smallest contacts of monovalent metals (gold) is carried by a single mode. For increasing contact size it is found that the transmission for the first mode goes to unity, before the second mode opens, then the second mode goes fully open, before the third opens, and so on. This property of atomic size contacts has been described as ``saturation of the channel transmission''\index{Saturation of channel transmission} \cite{ludoph99}. This property holds to very good approximation (a few percent) up to $G=1 G_0$, but deviations increase to 20\% admixture of the next channels at $G=4G_0$.

It is interesting to compare the positions of the maxima in the conductance histogram\index{Conductance histogram} and those for the minima in $\sigma_{GV}$ in Fig.~\ref{fig:2deriv}. It appears that these positions do not all coincide, which is most evident for the peak in the histogram at about $G=1.8 G_0$. The histograms give preferential conductance values, which may reflect a quantisation effect in the conductance as a function of contact diameter, but also a preference for forming contacts of certain effective diameters. Such preferential contact diameters may be expected based on the fact that the contact is only a few atoms in cross section, which limits the freedom for choosing the diameter. It appears that at least the peak at 1.8~$G_0$ in the histogram for gold arises from this atomic geometry effect. Although the shot noise and conductance fluctuation experiments both show that the conductance for gold contacts with $G\simeq 2 G_0$ is carried by two nearly perfectly transmitted modes, this conductance is not preferred, as evidenced by the conductance histogram. 

This observation amplifies the arguments given in the discussion of Fig.~\ref{fig:integrated} where it was argued that it is not appropriate to describe the conductance for gold as being quantised. There is no pronounced preference for conductances near multiples of $G_0$, with the exception of 1~$G_0$, which is due to a special property of gold contact in that it forms chains of atoms\index{Atomic chain} \index{Chain of atoms} at the last stages of contact elongation (see below). Of course, the conductance is a true quantum property in the sense that it is carried by only a few well-defined modes, but the transmission for these modes can add up to any value of the total conductance. Moreover, we identify a property for these contacts, called the saturation of channel transmission,\index{Saturation of channel transmission} which describes the tendency for the conductance modes to open one-by-one. 

For $s$, $s$-$p$ and $s$-$d$ metals the results for the subgap structure on the smallest contacts agree with the predictions of the tight binding model of Cuevas et al.\  under the assumption that the last plateau, before the jump to tunnelling, consists of a single atom contact. The agreement between the predicted number of channels and the number obtained from the fits of the current--voltage curves, thus confirm the assumption that the contacts consist of a single atom at the last stage of the contact elongation. The results on shot noise and conductance fluctuations for Al, where no quantum suppression was found, confirm the results by Scheer et al.\  that the conductance near $G=1\, G_0$ cannot be described by a single mode. We conclude that a saturation of channel transmission\index{Saturation of channel transmission} is not observed for Al, and is expected to be absent in all metals other than simple $s$-metals. Despite this apparent lack of a simple quantum mode structure, the conductance histogram\index{Conductance histogram} for aluminium\index{Al} shows pronounced peaks near the first three or four multiples of $G_0$ \cite{yanson97}, albeit considerably shifted from perfect integers. A natural interpretation for the first peak in the histogram, which is consistent with the findings above, would be that it arises from a reproducible last contact configuration of a single atom, with a conductance close to 1~$G_0$ and three channels involved in the conductance. In this sense, the histogram peak would arise purely as a result of atomic structure. In analogy, the other peaks are also expected to relate to atomic structure, although it is less evident what this structure should be. In the simplest approach one could imagine that the peaks correspond to 1, 2, 3 and 4 atoms in the contact cross section.

Further support for the model of valence orbitals as a basis for the conductance channels comes from the anomalous slope of the plateaux of the conductance for aluminium,\index{Al} where conductance increases as the contact is stretched ( Fig.~\ref{fig:aluminium}). Calculations of the conductance through a single atom, within the tight binding model discussed above, as a function of the bond distance of the atom with its neighbours, reproduce the anomalous dependence of conductance on distance \cite{cuevas98b}. For gold a nearly flat dependence is found, while for Pb the conductance decreases with increasing elongation, again in agreement with the observations. The anomalous slopes for Al are also found for larger contacts, and have been explained in terms of the stress-dependence of the electronic band structure \cite{sanchez}. It appears that this property of the band structure is conserved in contacts down to the atomic scale.  

\section{Chains of Atoms}
\index{Atomic chain}
\index{Chain of atoms}
All evidence shows that for a single atom contact for monovalent metals the current is carried by a single mode, with a transmission probability close to one. Guided by this knowledge, in experiments on gold\index{Au} Yanson et al.\  \cite{yanson98} discovered that during the contact breaking process the atoms in the contact form stable chains of single atoms, up to 7 atoms long. Independently, Ohnishi et al.\   \cite{ohnishi} discovered the formation of chains of gold atoms at room temperature in a combined STM and transmission electron microscope,\index{Transmission electron microscope} where an atomic strand could be directly seen in the images. Currently, the only material for which this chain formation is observed is gold. It is interesting that for silver the effect is much less pronounced and copper does not seem to show it at all.

\begin{figure}[!t]
\includegraphics[width=.8\textwidth]{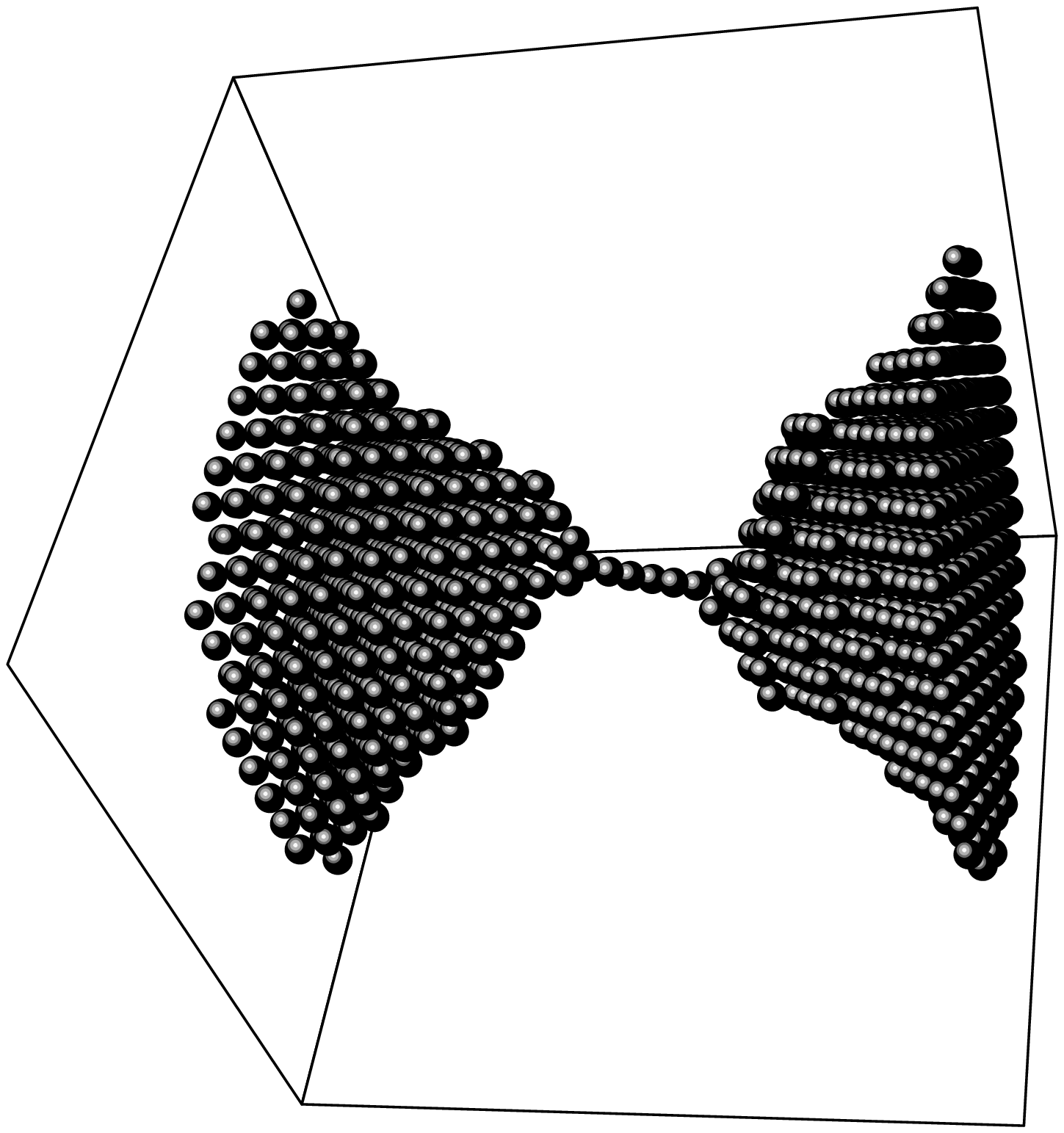}
\caption{Atomic configuration obtained during the last stages of breaking of a nanowire in a molecular dynamics simulation for gold, using a bath temperature of 12~K. (From \protect\cite{sorensen})}
\label{fig:chain}
\end{figure}

Some understanding of the underlying mechanism can be obtained from molecular dynamics simulations.\index{Molecular dynamics simulation} Already before the experimental observations, several groups had observed the spontaneous formation of chains of atoms in computer simulations of contact breaking \cite{sorensen,finbow,sutton}. Figure~\ref{fig:chain} shows the results obtained by S\o rensen et al.\  for gold. The authors caution that the interatomic potentials used in the simulation may not be reliable for this unusual configuration. However, from these studies we may learn more about the atomic configuration sequences producing the chains, and they form a good starting point for more advanced model calculations.

Such chains constitute the ultimate one-dimensional metallic nanowires. The current is carried by a single mode, with a transmission probability, which is somewhat below 1 due to back-scattering, but can be tuned to unity by adjusting the stress on the junction. The chains sustain enormous currents, up to 80~$\umu$A, due to the ballistic nature of the electron transport \cite{yanson98}. More work is needed to elucidate the mechanism of chain formation, what limits the length of the chains, and why it works best for gold. Further unusual atomic configurations may be found, according to model calculations by G\"ulseren et al.\   \cite{gulseren}, who show a series of ``weird wire''\index{Weird wires} structures depending on the number of atoms in the cross section of the wires and on the metallic element involved.

\section{Quantum Forces and Shell Structure in Alkali Nanowires}
\index{Quantum force}
\index{Shell structure}
\begin{figure}[!t]
\includegraphics[height=.8\textwidth,angle=270]{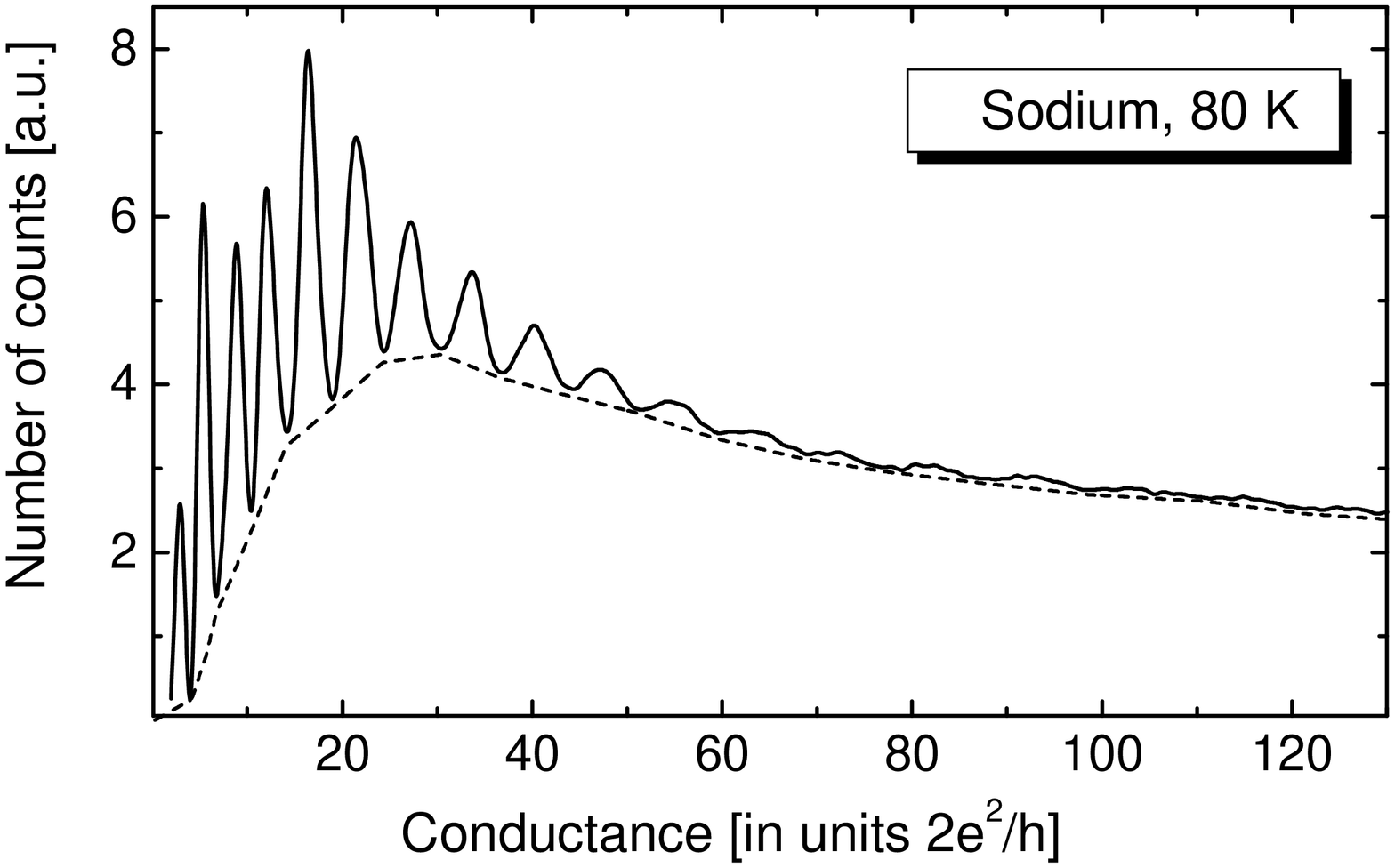}
\caption{Histogram of the number of times each conductance is observed versus the conductance in units of the conductance quantum, $G_0$, 
for sodium at $T$=80~K and bias voltage $V$=100 mV, constructed from over 10,000 individual scans. The smooth background ({\it dashed curve}) helps to bring out the smaller amplitude oscillations. (From \protect\cite{yanson99})}
\label{fig:NaShell}
\end{figure}

Apart from their role in determining the electronic transport properties of the contacts, the quantisation of the wave functions may also affect the energy of formation of the contact. In direct analogy with the effect of electronic shell closing on the formation energy of clusters \cite{de Heer}, one expects that specific contact diameters will be stabilised by the formation of quantum modes. 
Several groups have recently presented model calculations on this problem, mostly considering free electrons confined by a hard wall potential \cite{stafford97,jvr,yannouleas,blom,kassubek,hoppler}. Local density calculations taking the atomic structure into account have also appeared \cite{barnett,hakkinen,nakamura}. There appears to be a consensus on the magnitude of the force fluctuations which may result from this mechanism, which is of the order of 1~nN, and is comparable to the force jumps for the smallest contacts observed in Fig.~\ref{fig:rubio}. Some authors argue that the jumps observed in the conductance as a function of contact stretching should therefore be considered as being the result of the underlying electronic quantum modes. However, in the experiment many conductance steps are found which are much smaller, or much larger, than a conductance quantum, and all steps show similar mechanical and dynamic behaviour. In addition, the conductance steps in monovalent metals look similar to those in $sp$ or $sd$ metals, where such simple description definitely breaks down. Experimental methods will have to be developed to test a possible quantum-mode-based atomic force mechanism. One possibility is to measure the charge, or the work function of the nanowire, which is expected to fluctuate in unison with the force within this simple free electron gas picture \cite{jvr,kassubek}.

\begin{figure}[!t]
\includegraphics[width=.8\textwidth]{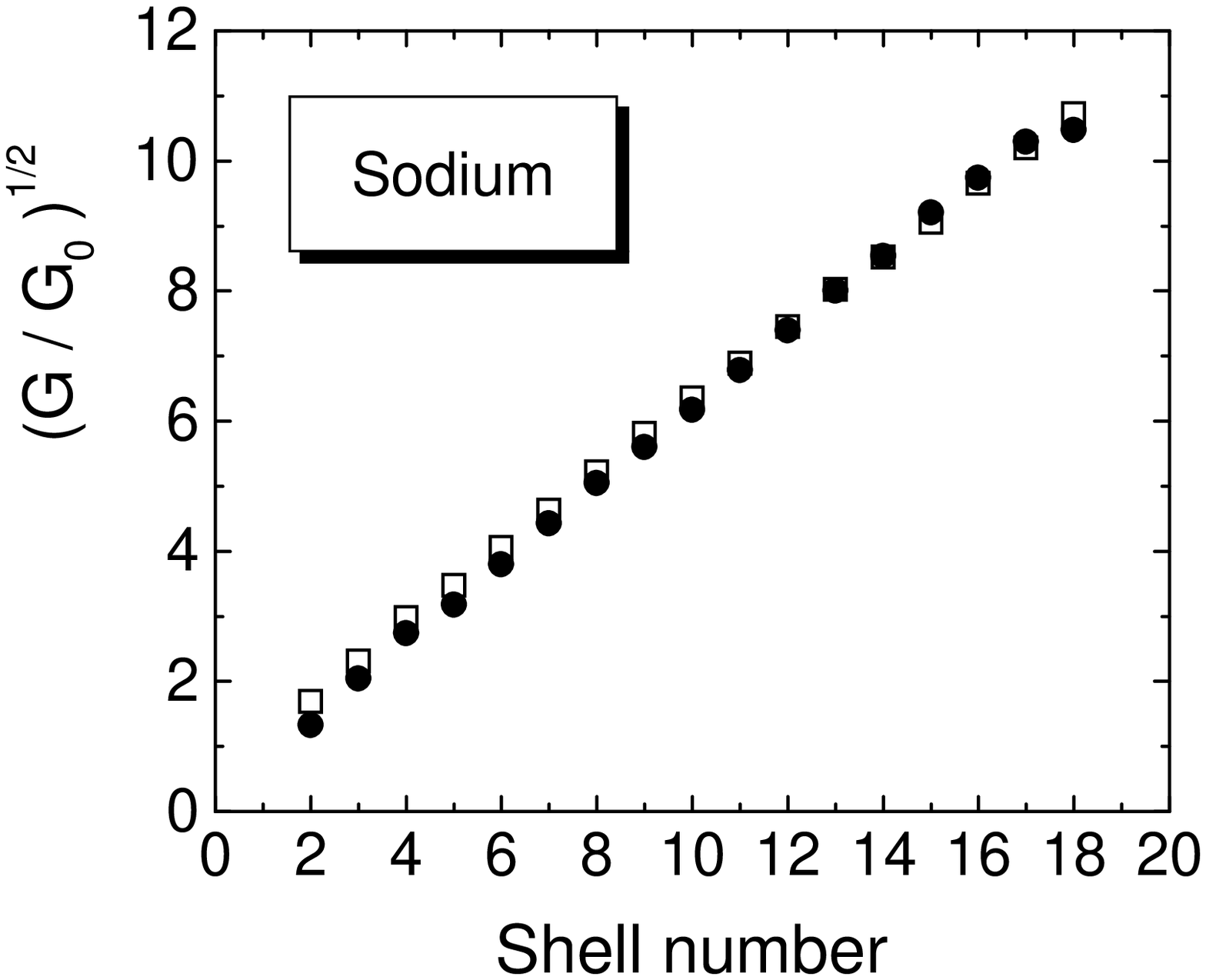}
\caption{Square root of the positions of the maxima in Fig.~\protect\ref{fig:NaShell}, $\sqrt{G_i/G_0}$, versus the shell number ({\it open squares}). The experimental peak positions are compared to the expected positions derived from the magic numbers of metallic clusters ({\it closed circles}). (From \protect\cite{yanson99})}
\label{fig:Magic}
\end{figure}

While the contribution of quantum modes to the force in the smallest contacts is still under debate, a newly observed phenomenon for larger contacts provides strong evidence for quantum force fluctuations.
In a recent study of conductance histograms for sodium,\index{Na} potassium\index{K} and lithium\index{Li} up to conductances much larger than shown in Fig.~\ref{fig:histK}, Yanson et al.\   \cite{yanson99} observed a large number of additional peaks (Fig.~\ref{fig:NaShell}). The peaks are not as sharp as the ones associated with conductance quantisation at low conductance (Fig.~\ref{fig:histK}) and cannot be identified with multiples of the conductance quantum. The peaks become more pronounced as the temperature is raised to about 80~K, and the position of the peaks is seen to be periodic in the square root of the conductance, as illustrated in Fig.~\ref{fig:Magic}. 

The interpretation of the phenomenon is based on  fluctuations in the density of states as a function of energy (or equivalently as a function of diameter) for a free electron gas inside a cylindrically symmetric wire. At the points where a new mode (the bottom of a 1-dimensional subband) crosses the Fermi energy, the density of states shows a $1/\sqrt{E-E_n}$ singularity. These singularities are smeared out by the finite length of the wire and would not have a very pronounced effect were they homogeneously distributed. However, the symmetry of the wire gives rise to a bunching of the singularities, which can be associated with the electronic shells in metal clusters. The resulting density of states fluctuations have been analysed by Stafford and coworkers \cite{stafford97,kassubek}, by Yannouleas et al.\  \cite{yannouleas} and by H\"{o}ppler and Zwerger \cite{hoppler}. The fluctuations in the density of states result in local minima in the total energy for the nanowire, and the stable wire diameters predicted from this model are in fairly good agreement with the observed periodic peak structure in the histograms \cite{yanson99}.

It was found that a direct comparison with cluster magic 
numbers\index{Magic number} is possible. One can calculate the effective cluster radius from the magic numbers for sodium clusters \cite{de Heer} $N_{m} = 8, 20,..., 1500$ using $R_{m} = r_{\rm S} N_{m}^{1/3}$, where $r_{\rm S}$ is the Wigner--Seitz radius of the atom. Then the conductance of a nanowire with these magic radii can be found using the semi-classical expression \cite{torres94}, $G_{m}=G_0(\pi R_{m}/\lambda_{\rm F})^2(1-\lambda_{\rm F}/\pi R_{m})$. The square root of the conductance values obtained in this way are plotted against the shell number $m$ and compared to the histogram peak positions in Fig.~\ref{fig:Magic}. A striking agreement is observed, which is believed to result from the fact that the dominant fluctuation terms in the density of states can be described in terms of the lowest order semi-classical trajectories inside the system \cite{balian,brack93,hoppler}. These trajectories (diametrical, triangular and square orbits inscribed inside the sphere and cylinder, respectively) are the same for clusters and nanowires. However, the agreement in Fig.~\ref{fig:Magic} is better than expected since the relative contribution of each of the trajectories for the two systems should be different. Further work is needed to clarify the details of the shell structure in nanowires.

\section{Discussion and Outlook}

With the developments of the last few years a microscopic understanding of atomic-scale electrical transport properties is beginning to grow. A coherent picture of conductance modes in a single atom derived from atomic valence orbitals is obtained, which finds strong support in the experimental observations for various metallic elements. For contacts of several atoms, monovalent metals demonstrate saturation of channel transmission,\index{Saturation of channel transmission} which is a very interesting experimental observation, but is not yet fully understood on the microscopic level.  Also, the nature of peaks in the histograms is still not fully resolved: quantisation plays a role but not exclusively. This is evident from the histograms for aluminium, in combination with the channel numbers obtained from subgap structure,\index{Subgap structure} \index{Superconducting contact} shot noise\index{Shot noise} and conductance fluctuations.\index{Conductance fluctuations} The peak at 5~$G_0$ in sodium is stronger than expected from a simple free electron nanowire model, and the peak below 2~$G_0$ in gold does not correspond to the position of the conductance quantum, even after correction for a series resistance, as was seen in  conductance fluctuation experiments. 

It seems clear that a full understanding of the structure in conductance histograms\index{Conductance histogram} requires a description of the mechanical evolution of the atomic structure of the contacts. It is possible that conditions such as temperature or bias voltage must be taken into account \cite{bratkovsky95}. Much has been learned already from molecular dynamics calculations,\index{Molecular dynamics simulation} but new experimental tools are probably essential. Promising developments come from two groups, which have recently shown transmission electron microscopy\index{Transmission electron microscope} images with atomic resolution of contacts that can be controlled in situ \cite{ohnishi,kondo,kizuka}. By simultaneous measurement of the conductance while observing a single atom contact in the electron micrograph, Ohnishi et al.\   clearly confirmed that a single atom gold contact has a conductance close to 1 $G_0$.

The question of the effect of the conductance modes on the force in metallic contacts will, without doubt, receive a lot of attention in the near future. A related question at present being considered is whether the increased density of states at specific wire diameters may be lifted by deformation of the shape of the wire cross section (in analogy to the Jahn--Teller effect) \cite{jvr} or by a spontaneous magnetisation \cite{zabala}. Many other open questions promise new and interesting developments in this area of physics. The dynamics of atomic structures, seen as two-level fluctuations, have not received sufficient attention. A related problem concerns the interaction with the electric current and the heating of the contact \cite{todorov98}. The observed chain formation for gold may lead to a deeper understanding of bonding forces on the atomic scale. We may be able to produce much longer atomic wires, which will open a new field for study of purely one-dimensional solid state physics.

\clearpage
\addcontentsline{toc}{section}{Index}
\flushbottom
\printindex

\end{document}